\documentclass{article}
\usepackage{graphicx} 
\usepackage{amsmath}
\usepackage{amsthm}
\newtheoremstyle{boldremark}
  {\topsep}   
  {\topsep}   
  {\itshape}  
  {}          
  {\bfseries} 
  {:}         
  {.5em}      
  {}          
\theoremstyle{boldremark}
\newtheorem*{remark}{Remark}
\usepackage{mathrsfs}
\usepackage{appendix}
\usepackage{array}
\usepackage{longtable}
\usepackage[authoryear,round]{natbib}
\usepackage{float} 
\usepackage{algorithm} 
\usepackage{algpseudocode}
\usepackage[hidelinks]{hyperref}
\usepackage{doi} 
\usepackage{booktabs}
\usepackage[table]{xcolor}
\usepackage[acronym]{glossaries}
\makenoidxglossaries
\usepackage[margin=1in]{geometry}
\usepackage{microtype}
\usepackage[section]{placeins}
\usepackage{tikz} 
\usetikzlibrary{arrows.meta}

\usepackage{xurl}

\newacronym{vwcl}{VWCL}{volume-weighted chain ladder}
\newacronym{oos}{OOS}{out-of-sample}

\ifdefined\assabuild
\usepackage{etoolbox}

\makeatletter

\newcounter{assa@level}
\newcommand{\assa@track}[2]{%
  \pretocmd{#1}{\setcounter{assa@level}{#2}}{}{%
    \GenericError{}{assa-numbering: could not patch \string#1}{}{}}}
\assa@track{\section}{1}
\assa@track{\subsection}{2}
\assa@track{\subsubsection}{3}

\newif\ifassa@number
\assa@numbertrue

\newcommand{\assa@exempt}[1]{\AtBeginEnvironment{#1}{\assa@numberfalse}}
\assa@exempt{figure}
\assa@exempt{table}
\assa@exempt{tabular}
\assa@exempt{longtable}
\assa@exempt{algorithm}
\assa@exempt{algorithmic}
\assa@exempt{tikzpicture}
\assa@exempt{remark}
\assa@exempt{abstract}
\assa@exempt{thebibliography}
\assa@exempt{theglossary}
\assa@exempt{description}

\pretocmd{\@hangfrom}{\assa@numberfalse}{}{%
  \GenericError{}{assa-numbering: could not patch \string\@hangfrom}{}{}}

\newcommand{\assa@print}[1]{\textbf{#1}\kern1em\relax}
\newcommand{\assa@parnumber}{%
  \ifassa@number
    \ifcase\value{assa@level}%
      \relax 
    \or \stepcounter{subsection}\assa@print{\thesubsection}%
    \or \stepcounter{subsubsection}\assa@print{\thesubsubsection}%
    \or \stepcounter{paragraph}\assa@print{\theparagraph}%
    \fi
  \fi
}
\AddToHook{para/begin}{\assa@parnumber}

\makeatother

\fi

\title{Supervising the Chain Ladder}

\author{
  Stephan Marais\thanks{Corresponding author.} \\
  \small Dynamo Analytics \\
  \small \texttt{stephan.marais@dyna-mo.com}
  \and
  James Grove \\
  \small Dynamo Analytics \\
  \small \texttt{james.grove@dyna-mo.com}
}
\date{October 2026}

\begin{document}

\maketitle

\begin{abstract}
The chain ladder's volume-weighted pattern minimises an explicit loss function, yet is rarely booked as such. Practitioners adjust the pattern and record the final adjusted ratios. This paper treats the chain ladder's pattern selection as a supervised-learning problem. Judgement on pattern adjustments becomes a framework of defined penalties and hyperparameters on the chain ladder's loss function, treated here as an objective function in machine learning. Data weights are generalised with a decay and a power parameter for recency and volume weighting. Benchmark shaping and smoothness enter through a reference penalty and Whittaker-Henderson smoothing. The assembled objective is strictly convex and minimised by a single linear system. Each hyperparameter becomes an interpretable adjustment in its own right, declarable by judgement and categorised as an experience or a prospective adjustment. Experience adjustments can be set more objectively by a proposed training loop and a reserve validation score on held-out calendar diagonals. Further hyperparameter-based adjustments are written as almost-everywhere differentiable penalties that re-time or reshape the pattern. A worked example carries one real Schedule P triangle through an incurred and then a paid training stage, demonstrating the workflow.
\end{abstract}

\vspace{1em}

\section{Introduction}\label{sec:introduction}

Actuaries describe real-world processes with elegant mathematics in the form of models. Assumptions are stated, and estimates derived from the data by robust mechanisms. Unfortunately, the mathematics diverges from real-world events, and the resulting disorder forces actuaries to make tactical, often messy, adjustments to their models. This paper approaches these adjustments of the model with the mathematical rigour they deserve, applied to the chain ladder.

Claims development carries nuances and unknowns that the mathematics does not express: a strengthening of case reserves, a backlog in the courts, a book changing shape. Reserving practice bridges the gap pragmatically, adjusting the model by hand until it agrees with what the actuary knows. The adjustments are sensible, but they live outside the model. They end up subjective, they differ between practitioners holding the same view, and the record keeps the adjusted figures rather than the true objectives behind the adjustments.

The chain ladder is a classic example of this gap. It predicts outstanding claims through successive development ratios. \citet{ref:Mack1993} showed that it is a statistical model: the \gls{vwcl} pattern is the exact minimiser of an explicit volume-weighted loss function (Section~\ref{sec:fitted-model}). Yet the fitted pattern is rarely booked as it stands. The actuary adjusts it through the typical practices listed in Section~\ref{sec:by-hand}, and each adjustment changes the booked reserve while only the selected ratios are recorded. The rigour ends exactly where the adjustments begin.

Can the actuary's manual pattern adjustments be written as declared terms of the chain ladder's own objective function, without giving up the statistical model that makes it rigorous?

This paper argues that they can. The actuary's views of how the chain ladder should fit the data for its future estimates are written as generalised loss weights and penalty terms. These turn the loss function into a penalised objective function. The chain ladder is trained as a supervised-learning model, and the actuary still supervises what it learns.

Judgement is relocated, not removed. The many ratio edits, difficult to document, become a small set of declared hyperparameters. No prior work has framed the practitioner's pattern adjustment toolkit as additive penalty terms in the chain ladder's objective function. Penalty terms of this kind are common in supervised-learning problems.

A factor selected by judgement is already an implicit choice of model (Section~\ref{sec:by-hand}), and the penalty terms introduced in this paper are not new to actuarial work. What this paper contributes is the assembly of the objective function and the framing of training the chain ladder model itself as a supervised-learning problem. This direction also differs from most machine learning in the reserving literature, which explores the use of flexible models, such as neural networks, in triangle reserving \citep{ref:deeptriangle}. This approach can lead to performance gains, but introduces challenges for model transparency and interpretability \citep{ref:AIinActSci}. This paper moves the other way, bringing the supervised-learning discipline to a model with which practitioners are familiar.

\citet{ref:CDRScore} showed how reserving models can be selected in a supervised-learning framework. They score existing methods, with adjustments such as using only the most recent accident years or excluding an extreme factor, on held-out diagonals. Those adjustments filter the data that the chain ladder averages into its development ratios. The penalties in this paper act instead inside the objective function that the development ratios minimise. They cover shape adjustments, smoothing, benchmarking and re-timing. The paper also demonstrates how the objective function's hyperparameters can be trained in a loop similar to the framework in \citet{ref:CDRScore}.

This paper has the following structure. The chain ladder is reframed as a model fitted under a declared loss function, as necessary background (Section~\ref{sec:fitted-model}). The different kinds of manual adjustment described in Section~\ref{sec:by-hand} are encoded as declarable objective-function terms, derived in Section~\ref{sec:building}. The framework is extended with three hyperparameter-based pattern adjustments (Section~\ref{sec:adjustments}). A reserve validation score and a training loop that selects the hyperparameters are introduced and demonstrated on a real Schedule P dataset (Section~\ref{sec:model-training}). Section~\ref{sec:discussion} sets out the limitations, and Section~\ref{sec:conclusion} concludes.

\section{The chain ladder as a fitted model}\label{sec:fitted-model}

In macro reserving the data are the cumulative claim amounts $C_{i,j}$, indexed by accident period $i$ and development period $j$. For simplicity both periods share one periodicity, both monthly or both annual, so the observed data form a triangle. At the valuation date, $C_{i,j}$ is observed for $i + j \le I + 1$, where $I$ is the number of accident periods. The calendar period of a claim is $k = i + j - 1$, so the most recently observed diagonal is $k = I$. We assume the book is fully run off by development period $I$, so no tail factor is required. The triangle may be paid or incurred. Nothing in the paper depends on which, and the notation does not distinguish them. The reserving goal is to predict the unobserved future values so that the actuary knows how much to hold for the claims still to be paid. Table~\ref{tab:notation} collects the notation used throughout; each quantity is defined where it first arises, in the section referenced.

\begin{table}[!htbp]
    \centering
    \begin{tabular}{ll}
        \toprule
        \rowcolor{gray!15} Symbol & Meaning \\
        \midrule
        $C_{i,j}$ & cumulative claims for accident period $i$ at development period $j$ \\
        $I$ & number of accident periods; the triangle is square, $j = 1, \ldots, I$ \\
        $k = i + j - 1$ & calendar period of claim $(i,j)$; the latest observed diagonal is $k = I$ \\
        $f_{i,j} = C_{i,j+1}/C_{i,j}$ & individual development ratio, observed for $i + j \le I$ \\
        $f_j$, $\boldsymbol{f}$ & the development ratios being fitted, $j = 1, \ldots, n$ \\
        $n = I - 1$ & number of development ratios \\
        $\sigma_j^2$ & development variance parameter of period $j$ (Section~\ref{sec:validation-score}) \\
        $w_{i,j}$ & loss weight on $f_{i,j}$ (Section~\ref{sec:decay}) \\
        $\rho, \gamma$ & decay and power parameters of the loss weights (Section~\ref{sec:decay}) \\
        $W_j = \sum_{i=1}^{I-j} w_{i,j}$ & total weight at development period $j$ (Section~\ref{sec:reference}) \\
        $\operatorname{Err}$ & the weighted empirical loss (Sections~\ref{sec:by-hand} and~\ref{sec:decay}) \\
        $J_p$, $\alpha_p$ & penalty term $p$ and its weight in the objective (Section~\ref{sec:by-hand}) \\
        $\mathscr{L}$ & the objective function, the weighted loss plus the penalties (Section~\ref{sec:by-hand}) \\
        $\dot f_j$ & the reference (a priori) pattern (Section~\ref{sec:reference}) \\
        $\hat f_j$, $\hat{\boldsymbol{f}}$ & the fitted pattern, the objective's minimiser at given hyperparameters (Section~\ref{sec:objective}) \\
        $\hat f_j^{\,\text{CL}}$ & the unpenalised \gls{vwcl} pattern \\
        $\bar f_j$ & the weighted empirical solution (Section~\ref{sec:decay}) \\
        $\alpha_{\text{ref}}, \alpha_{\text{smooth}}$ & penalty weights (Sections~\ref{sec:reference} and~\ref{sec:smoothness}) \\
        $\alpha_{\text{shift}}, \alpha_{\text{rot}}$ & shift and rotation penalty weights (Sections~\ref{sec:shift} and~\ref{sec:rotate}) \\
        $d$ & drift strength of the pattern drift adjustment (Section~\ref{sec:drift}) \\
        $\rho_{\text{score}}$ & decay weight aggregating the validation splits' scores (Section~\ref{sec:validation-score}) \\
        \bottomrule
    \end{tabular}
    \caption{Notation. Boldface denotes the vector over development periods, and a hat denotes a fitted or projected quantity.}
    \label{tab:notation}
\end{table}

\subsection{The volume-weighted chain ladder}

The most common way to fill in the future triangle is the chain ladder, which assumes a multiplicative relationship between successive cumulative amounts: $\hat C_{i,j+1} = C_{i,j} \cdot f_j$. The \gls{vwcl} estimates the development ratios as
\[
    \hat f_j^{\,\text{CL}} = \frac{\sum_{i=1}^{I-j} C_{i,j+1}}{\sum_{i=1}^{I-j} C_{i,j}},
\]
which is intuitive when the cumulative amounts are viewed in their triangle format: the sum of one column divided by the sum of the column before it, over the accident periods where both are observed.

The same estimator can be read as a statement about the \textit{individual development ratios} $f_{i,j} = C_{i,j+1}/C_{i,j}$, which give the identity $C_{i,j} = C_{i,1} \prod_{m=1}^{j-1} f_{i,m}$. Each $f_{i,j}$ is one accident period's observation of the development from $j$ to $j+1$, and the \gls{vwcl} is their weighted average, weighted by the preceding cumulative amount:
\begin{align}\label{eq:cl-as-weighted-average}
    \hat f_j^{\,\text{CL}} = \frac{\sum_{i=1}^{I-j} C_{i,j+1}}{\sum_{i=1}^{I-j} C_{i,j}}
    = \frac{\sum_{i=1}^{I-j} \frac{C_{i,j+1}}{C_{i,j}} \cdot C_{i,j}}{\sum_{i=1}^{I-j} C_{i,j}}
    = \frac{\sum_{i=1}^{I-j} f_{i,j} \cdot C_{i,j}}{\sum_{i=1}^{I-j} C_{i,j}}.
\end{align}
This reading is used throughout the paper: the fitted pattern is a weighted average of observed ratios, and the weights are a modelling choice.

\subsection{The chain ladder as a statistical model}

\citet{ref:Mack1993} expresses the chain ladder as a statistical model, treating the development ratios as parameters to be estimated under explicit assumptions on the error. \citet{ref:mack1994} expands that treatment and supplies the regression reading used below. Writing the one-step relationship with an additive error,
\[
    C_{i,j+1} = C_{i,j} \cdot f_j + \epsilon_{i,j},
\]
where the $\epsilon_{i,j}$ are independent with $E(\epsilon_{i,j}) = 0$, a natural requirement is that $\boldsymbol{f}$ minimise the sum of squared errors. The errors are not identically distributed, however: \citet{ref:Mack1993} assumed that the conditional variance is proportional to the preceding volume,
\[
    \operatorname{Var}(C_{i,j+1} \mid C_{i,1}, \ldots, C_{i,j}) = C_{i,j} \cdot \sigma_j^2,
\]
where $\sigma_j^2$ is the development variance parameter of period $j$, constant across accident periods. Weighting each squared error by the reciprocal of its variance - the standard generalised least-squares prescription - gives the volume-weighted squared-error loss
\begin{align}\label{eq:vw-loss}
    L(\boldsymbol{f})
    = \sum_{j=1}^{n} \sum_{i=1}^{I-j} \frac{1}{C_{i,j}} \left(C_{i,j+1} - C_{i,j} \cdot f_j\right)^2
    = \sum_{j=1}^{n} \sum_{i=1}^{I-j} C_{i,j} \left(f_{i,j} - f_j\right)^2.
\end{align}
Two features of Equation~\eqref{eq:vw-loss} carry through the whole paper. First, it can be written entirely in terms of the development ratios and the individual development ratios, which is the form the penalty terms of later sections attach to. Second, the ratios separate: each $f_j$ appears only in its own inner sum, so the loss is minimised ratio by ratio. Setting the derivative with respect to $f_j$ to zero shows that the minimiser is exactly the \gls{vwcl} solution of Equation~\eqref{eq:cl-as-weighted-average} \citep{ref:mack1994}. The \gls{vwcl} is therefore not merely a convenient recipe. It is the fitted model under a declared loss function, and fitting it can be treated as a supervised-learning problem.

\subsection{The triangle data}\label{sec:running-triangle}

The mechanisms of Sections~\ref{sec:decay}--\ref{sec:smoothness} and the adjustments of Section~\ref{sec:adjustments} are all shown on one synthetic illustration triangle. It is a monthly paid triangle of 120 accident months, generated from a simulated claims process. Its generator deliberately builds in characteristics that actuaries often adjust for by hand in a chain ladder fit. The \gls{vwcl} pattern is visibly rough. Calendar shocks in claims inflation move recent experience away from the older accident months. The data run to 120 development months. Where an annual version of the triangle is clearer, the same book is aggregated to a triangle of ten accident years and ten development periods. The demonstration on real data is deferred to the worked example of Section~\ref{sec:example}.

\begin{figure}[!htbp]
  \centering
  \includegraphics[width=0.75\textwidth]{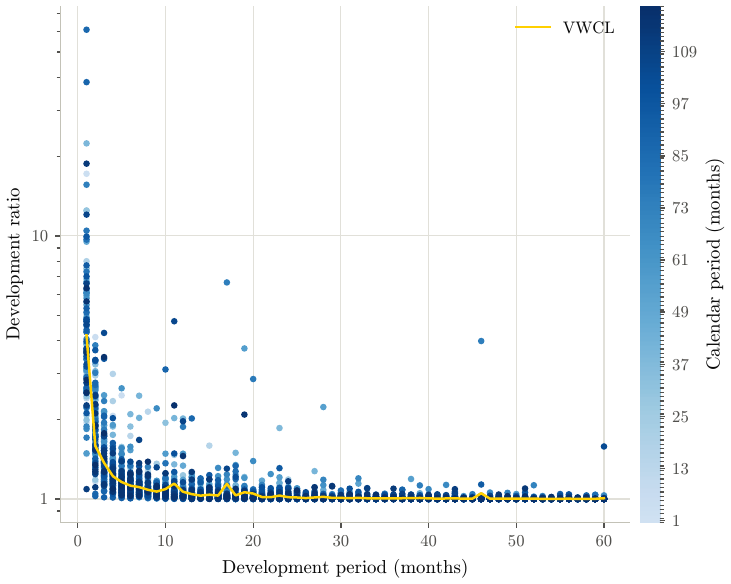}
  \caption{The triangle data: individual development ratios (points, log scale) by development month over the first 60 of 120 development months, coloured by calendar period, with the unpenalised \gls{vwcl} pattern (line) overlaid. The simulated characteristics are visible: a rough pattern and calendar shocks.}
  \label{fig:running-triangle}
\end{figure}

Figure~\ref{fig:running-triangle} shows the first 60 development months of the triangle's individual development ratios, with the unpenalised \gls{vwcl} pattern of Equation~\eqref{eq:cl-as-weighted-average} overlaid. The fitted pattern inherits the characteristics in the data: it is jagged from period to period, and it averages over the calendar year differences rather than tracking them. An actuary presented with this fit would be reluctant to book it as is; they would first adjust it by hand. The next section describes those adjustments. The remainder of the paper shows how the gap between this fit and a defensible pattern can be closed without manual adjustments.

\section{Formalising what actuaries do by hand}\label{sec:by-hand}

Section~\ref{sec:fitted-model} established that the \gls{vwcl} pattern is the minimiser of a declared loss function. In practice the fitted pattern is rarely booked unchanged. A reserving actuary reviewing the fitted development ratios will routinely: exclude an individual ratio that reflects a one-off event, a data error or an outlier; give more weight to recent periods when the book is changing over time; smooth a pattern whose period-to-period jaggedness is seen as noise; pull the pattern towards a familiar benchmark for the class, from industry data or an a priori view; check the paid projection against the incurred projection run beside it, reconciling the two views before booking either; and shift the pattern earlier or later when settlement behaviour is known to have changed. Every one of these is standard practice, and every one of them changes the booked reserve.

Each of these adjustments encodes an \textit{objective} about the future that goes beyond fitting the past data. Excluding a ratio asserts that the future does not resemble the distribution of the data with that observation included. Weighting recency asserts that the data-generating process is drifting and that recent periods are more representative of the future. Smoothing asserts that the true development pattern is smooth even where the observed one is not. Benchmarking asserts that information from outside the triangle should constrain the fit. Reconciling paid against incurred asserts that the two views measure the same underlying claims, so each carries information about the other's future development. Manipulating the shape of the pattern asserts a known or expected change in settlement behaviour that the triangle history does not yet show. The actuary performing these adjustments is correcting the model's fit towards what they believe to be the true underlying distribution of the claims data.

By making these adjustments, the actuary is (manually) training the model towards objectives the loss function does not contain. \citet{ref:BardisMajidiMurphy2013} show that, under certain restrictions, an actuary's manually selected factor is the best linear unbiased estimate for a member of a continuously indexed family of chain ladder models. A selection made by judgement is therefore already an implicit model choice and can be seen as training the model.

We divide these adjustments into two categories:

An \textbf{experience adjustment} changes how the observed claims experience is read. It asserts that the raw triangle, as a sample of the insurance book, can misrepresent the true mechanism that generates it. The adjustment aims to fix that misrepresentation: an unrepresentative ratio, noise in the data, a weighting that dilutes an underlying trend.

A \textbf{prospective adjustment} is anticipatory. It asserts that the future will differ from the experience, however well that experience is read. Examples are a settlement pattern known to be shortening and a new industry trend the book is expected to follow, before the data can show either.

The distinction between these adjustments determines how each kind's weight can be chosen. An experience adjustment's effect can be observed in the historical data, so the supervised-learning machinery of Section~\ref{sec:training} can train its weight.
A prospective adjustment asserts precisely what no experience data can confirm, so its weight must be declared by the practitioner and cannot be learned from the data.

These manual adjustments can be reframed as penalties to the loss function in Equation~\eqref{eq:vw-loss} through two mechanisms. The first is \textit{reweighting the data}: deciding how much to trust each data point. Exclusions and recency weighting belong here. They act \textit{inside} the empirical loss, through the weights it assigns to the individual development ratios. The second is \textit{penalising the shape}: declaring what the fitted pattern should look like, independently of any single data point. Smoothness, benchmark shaping and the pattern shape adjustment belong here. They act as terms \textit{added to} the loss. The paid-incurred reconciliation is the one adjustment the framework carries only in part. A reference pattern derived from the companion view can be brought in through benchmark shaping (Section~\ref{sec:reference}). The two mechanisms combine into one objective function,
\begin{align}\label{eq:general-form}
    \mathscr{L}(\boldsymbol{f}) = \operatorname{Err}\!\left(\boldsymbol{f};\, w(\rho, \gamma)\right)
    + \sum_{p \in \mathcal{P}} \alpha_p \, J_p(\boldsymbol{f}),
\end{align}
where $\operatorname{Err}$ is the empirical loss under weights governed by a decay parameter $\rho$ and a power parameter $\gamma$, $\mathcal{P}$ is the declared set of penalties, each $J_p$ is a penalty expressing one declared objective about the shape of the pattern, and each $\alpha_p \ge 0$ sets how much that objective matters relative to fit. The fitted pattern becomes the minimiser of the objective $\mathscr{L}$ instead of Equation~\eqref{eq:vw-loss}.

The empirical term deliberately carries no weight of its own. Scaling the whole objective by a constant does not move its minimiser, so only the penalty weights \textit{relative to} the empirical loss matter. Fixing the empirical term's weight at one therefore removes a redundant dimension from the hyperparameter search of Section~\ref{sec:training}, without restricting the family of solutions.

Equation~\eqref{eq:general-form} structures the remainder of the paper: Section~\ref{sec:building} builds the objective term by term, and Section~\ref{sec:model-training} trains its hyperparameters.

\section{Building the objective}\label{sec:building}

Equation~\eqref{eq:general-form} named two mechanisms - reweighting the data and penalising the shape - and Section~\ref{sec:by-hand} mapped the manual adjustments onto them. This section builds the objective function term by term: the generalised loss weights, the penalty terms derived from the practices they replace, and the assembled objective and its solution. Every quantity is computed from the fitted triangle's data and the declared inputs alone, so the objective is available on the exact data used by the chain ladder itself.

Regularised estimation is not itself new to reserving. \citet{ref:JeongChangValdez2021} penalise development-factor estimation with a non-convex penalty to stabilise the estimator. \citet{ref:Venter2018} shrinks the parameters of over-parameterised reserving models by regularisation, as an alternative to the curves actuaries fit by hand. \citet{ref:McGuireTaylorMiller2021} automate reserving with the lasso, selecting the model by cross-validation. In each, the regularisation serves stability or automation. Here it is the language in which the practitioner's objectives are declared, term by term.

\subsection{Weighting data: exclusions and recency}\label{sec:decay}

The first mechanism of Section~\ref{sec:by-hand} is deciding how much to trust each data point. In practice this takes two familiar forms. An actuary excludes an individual ratio - a large loss settling unusually, a data correction, a period distorted by an event - by simply leaving it out of the average. When the book is changing, through growth, business mix or process change, the actuary leans more on recent diagonals than on older ones. These are experience adjustments as defined in Section~\ref{sec:by-hand}.

Both exclusions and recency become declarable by generalising the weights of the empirical loss. Replace the volume weights $C_{i,j}$ of Equation~\eqref{eq:vw-loss} with
\begin{align}\label{eq:weighted-loss}
    \operatorname{Err}(\boldsymbol{f})
    = \sum_{j=1}^{n} \sum_{i=1}^{I-j} w_{i,j} \left(f_{i,j} - f_j\right)^2,
    \qquad
    w_{i,j} = \rho^{\,I-i-j} \, C_{i,j}^{\gamma}.
\end{align}
The decay parameter $\rho \in (0, 1]$ sets the severity of the exponential decay, so that older data receive less weight than recent data. Exponential decay weights have a long precedent in reserving. \citet{ref:Gluck1997} introduced them to the Cape Cod method, weighting each accident year by its distance from the year being estimated. \citet{ref:CDRScore} list the same device as an option for weighting individual development ratios. The exponent $I - i - j$ is zero on the most recent diagonal, leaving it undecayed, and increases by one for each calendar period further back. The power parameter $\gamma$ determines how much claim size affects an individual development ratio's influence on the fitted ratio $\hat f_j$.

The minimiser remains a per-ratio weighted average of the individual development ratios,
\begin{align}\label{eq:weighted-solution}
    \hat f_j = \frac{\sum_{i=1}^{I-j} w_{i,j} \, f_{i,j}}{\sum_{i=1}^{I-j} w_{i,j}}
    \;=:\; \bar f_j,
\end{align}
the weighted empirical solution and the anchor that the penalties of Sections~\ref{sec:reference} and~\ref{sec:smoothness} pull against.

\begin{remark}
The familiar estimators for $f_j$ are special cases of this generic formulation. At $\rho = 1$ and $\gamma = 1$, Equation~\eqref{eq:weighted-solution} is the \gls{vwcl}. At $\gamma = 0$ it is the simple average of the individual development ratios, and at $\gamma = 2$ it minimises the unweighted squared error of the one-step predictions $C_{i,j} f_j$ \citep{ref:ClaimWeighting}.
\end{remark}

A data exclusion sets $w_{i,j} = 0$ for a ratio the actuary is concerned about. Exclusions, even if applied algorithmically or by rule, limit the objectivity of the model. The exclusions approach implies knowledge the actuary holds about the data themselves, or a reading of their distribution. Two practitioners can hold the same view of the data, yet apply exclusions differently. By contrast, the decay and volume weightings are governed by two tunable parameters, $\rho$ and $\gamma$.

\begin{figure}[!htbp]
  \centering
  \includegraphics[width=0.95\textwidth]{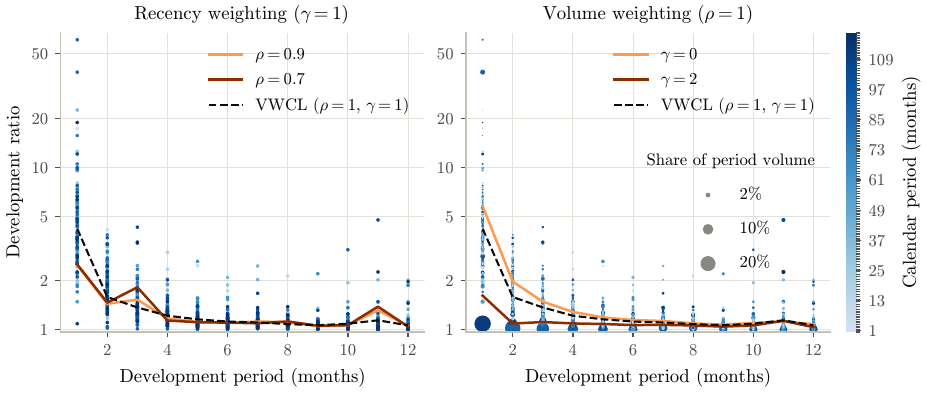}
  \caption{Weighting the data on the synthetic illustration triangle. The points are the individual development ratios $f_{i,j}$ on a logarithmic vertical axis, shaded by calendar period and darkest on the newest diagonals. On the right each point's area is its share of the total volume $\sum_i C_{i,j}$ at that development period, so the area it covers is the weight $\gamma = 1$ gives it.}
  \label{fig:decay}
\end{figure}

Figure~\ref{fig:decay} shows the mechanics of this weighting. Both panels shade the points by calendar period, darkest on the newest diagonals. Recency weighting ($\rho < 1$) pulls the fit towards the dark recent experience that the unweighted fit averages over. In the right panel each point's area is its share of the total volume at that development period. A point therefore covers the weight it carries at $\gamma = 1$. Raising the power parameter moves the fit through the special cases $\gamma = 0$, $1$ and $2$, trusting the large observations progressively more as it rises.

\FloatBarrier

This weighting for individual development ratios is not without precedent. For example, software such as the \texttt{chainladder-python} package expose the averaging method and the number of recent diagonals, \texttt{n\_periods}, as hyperparameters of its development estimator \citep{ref:BogaardtDanHsu2024}. That window is a discrete form of the recency weighting that $\rho$ makes continuous and declarable. \citet{ref:ClaimWeighting} defines the development-factor estimator with arbitrary per-observation weights $w_{i,j} \in [0,1]$ for downweighting outlying ratios. \citet{ref:Murphy1994} frames the development process as a least-squares regression and derives best linear unbiased estimates for the common link-ratio estimators. \citet{ref:BarnettZehnwirth2000} treat link-ratio techniques as weighted regressions and test their assumptions, as does \citet{ref:Venter1998}. Weighting also connects to robustness. \citet{ref:AvanziLavenderTaylor2023} derive impact functions that quantify the effect of each individual observation on chain ladder reserves and their errors. Robust reserving techniques limit such influence by detecting and adjusting outliers \citep{ref:VerdonckVanWouweDhaene2009}.
The intent differs, however. Robust estimation lets the data decide what to distrust. Here every downweighting is declared: an exclusion is named, and the recency and volume weightings are stated dimensions of trust that the validation loop of Section~\ref{sec:training} trains.

\subsection{Regression towards a benchmark}\label{sec:reference}

Section~\ref{sec:by-hand} describes adjustments based on a prior view of what the pattern should look like. When the triangle is thin, volatile or not far developed, the actuary pulls the selected ratios towards a reference pattern for the class: an industry pattern, a pattern from a larger related book, or an a priori view. Bringing a view from outside the triangle into the estimate is the essence of the Bornhuetter-Ferguson method \citep{ref:BornhuetterFerguson1972}, which trades the triangle's own experience against an external expectation. However, applying this Bayesian approach by hand directly on the pattern itself leaves no record of the adjustment in the model.

\subsubsection{The reference-pattern penalty}

This benchmark view can be expressed as a penalty term.

Let $\dot f_j$ denote the reference pattern. Then define the penalty as the weighted squared difference between the fitted and the reference pattern,
\begin{align}\label{eq:reference-penalty}
    J_{\text{ref}}(\boldsymbol{f}) = \sum_{j=1}^{n} W_j \left(f_j - \dot f_j\right)^2,
    \qquad
    W_j = \sum_{i=1}^{I-j} w_{i,j},
\end{align}
where $W_j$ is the total empirical-loss weight at development period $j$.

The choice of $W_j$ as the weighting inside the penalty is a design rule that recurs in Section~\ref{sec:smoothness}. The penalty carries the same weights as the empirical loss, so the effective weight is dictated by the hyperparameter $\alpha_{\text{ref}}$. Adding $\alpha_{\text{ref}} \, J_{\text{ref}}$ to the weighted loss of Equation~\eqref{eq:weighted-loss} keeps the objective separable per ratio, and minimising it yields (Appendix~\ref{ap:matrix})
\begin{align}\label{eq:credibility-blend}
    \hat f_j = \frac{\bar f_j + \alpha_{\text{ref}} \, \dot f_j}{1 + \alpha_{\text{ref}}},
\end{align}
a one-line credibility blend of the empirical solution and the reference pattern, with credibility $1/(1 + \alpha_{\text{ref}})$ on the data.

Three cases help the interpretation: $\alpha_{\text{ref}} = 0$ returns the empirical fit, $\alpha_{\text{ref}} = 1$ returns the average of the two, and $\alpha_{\text{ref}} \to \infty$ returns the benchmark. This is the exact statistical analogue of ridge regression shrinking coefficients towards a target \citep{ref:elementsSLT}. It also has a direct actuarial precedent. \citet{ref:GislerWuthrich2008} derive the chain ladder factor in a Bayesian set-up as precisely a credibility-weighted average between the triangle's own factor and an a priori expected value from expert opinion or market experience.

\begin{figure}[!htbp]
  \centering
  \includegraphics[width=0.75\textwidth]{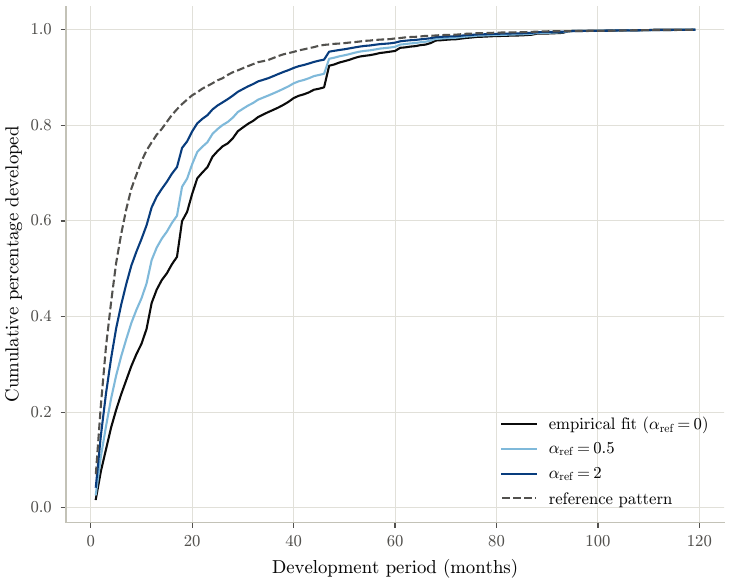}
  \caption{The reference-pattern penalty on the synthetic illustration triangle. As $\alpha_{\text{ref}}$ increases the fitted pattern (lines) moves from the weighted empirical solution towards the reference pattern (dashed), applying credibility $1/(1+\alpha_{\text{ref}})$ to the data at every development period.}
  \label{fig:reference}
\end{figure}

Figure~\ref{fig:reference} shows the penalty acting on the synthetic illustration triangle. Increasing $\alpha_{\text{ref}}$ moves the jagged empirical fit towards the reference pattern.

\subsubsection{Scaling to incorporate companion data}\label{sec:companion}

The reference pattern is not limited to external benchmarks but can be derived from other reserve projections already available. Suppose a chain ladder model is fitted on a paid triangle, with an incurred chain ladder already fitted beside it on the incurred data $C^I_{i,j}$. The incurred fit can then supply a paid reference pattern.

Reserving models draw on triangle data of many types, such as paid and incurred amounts on various bases, reported or settled claim counts, and exposure measures.
When projecting from a specific triangle, for example the paid triangle, we refer to the other data not used by the chain ladder as \textit{companion data}; models fitted on companion data produce companion projections or patterns.

For example, a paid reference pattern can be derived from the incurred triangle $C^I_{i,j}$ using a declared paid-to-incurred trajectory $q^*_j$, the share of the incurred amount expected to be paid by development period $j$. It converts the incurred view's volume-weighted development $\bar f^I_j = \sum_i C^I_{i,j+1} / \sum_i C^I_{i,j}$ into a paid reference, $\dot f_j = (q^*_{j+1}/q^*_j) \, \bar f^I_j$. A declared burning-cost pattern $b^*_j$ - cumulative claims expected per unit of exposure at each development period, the expected-loss view of the Cape Cod tradition \citep{ref:Stanard1985} - gives $\dot f_j = b^*_{j+1}/b^*_j$. A reported-count triangle $N_{i,j}$ with a declared average cost per reported claim $c^*_j$ gives $\dot f_j = (c^*_{j+1}/c^*_j) \, \bar f^N_j$, with $\bar f^N_j$ the count triangle's volume-weighted development.

\begin{remark}
Calibrating a companion reference pattern directly from the raw data, using the \gls{vwcl}, collapses it exactly to the fitted triangle's own volume-weighted ratios. It is the companion projection's training information that carries across in the reference pattern. Section~\ref{sec:example} demonstrates this.
\end{remark}

A companion reference inherits its companion's defects along with its information. The declared weight $\alpha_{\text{ref}}$ is then also a statement about the quality of the companion data.

\subsubsection{Alternative weighting strategies}

Whatever the source of $\dot f_j$, the weighting inside the penalty has so far been fixed at $W_j$, and that design rule carries a cost. Equation~\eqref{eq:credibility-blend} applies the same credibility at every development period. On the other hand, \citet{ref:GislerWuthrich2008} vary the credibility with volume, so their estimator leans harder on the collective where the data are thin. Recovering that behaviour costs nothing structurally. Replacing $W_j$ in Equation~\eqref{eq:reference-penalty} with a free diagonal weight $V_j > 0$ leaves the objective separable per ratio, and minimising it gives
\begin{align}\label{eq:credibility-blend-general}
    \hat f_j = a_j \, \bar f_j + \left(1 - a_j\right) \dot f_j,
    \qquad
    a_j = \frac{W_j}{W_j + \alpha_{\text{ref}} \, V_j},
\end{align}
still a weighted average of the two, but with the credibility $a_j$ now free to vary across development periods. Only the ratio $V_j / W_j$ enters the solution, and $V_j = W_j$ returns Equation~\eqref{eq:credibility-blend}.

Two logical choices are to hold $V_j$ constant, or to set it to a development-period variance estimate $\hat\sigma_j^2$, for example the formula of \citet{ref:Mack1993} estimated from the \gls{vwcl} ratios. At $\rho = 1$ and $\gamma = 1$ both land on the credibility weight of \citet{ref:GislerWuthrich2008}, with $\alpha_{\text{ref}}$ playing different roles in the within-group and between-group ratio. The main limitation is that $\alpha_{\text{ref}}$ is constant across development periods, which is unlikely to hold under their assumptions.

This paper adopts a constant $\alpha_{\text{ref}}$ and the consistent weighting $V_j = W_j$ independent of $i$, chosen for interpretability and for the direct comparison it allows between the competing objectives. A more dynamic weighting is left as future research.

\FloatBarrier

\subsection{Smoothing the pattern: the smoothness penalty}\label{sec:smoothness}

For the empirical loss in Equation~\eqref{eq:weighted-loss}, and more generally for any loss function that separates by development ratio, the fitted ratios are estimated independently. Consequently, nothing in the optimisation problem encourages neighbouring estimates to be similar - or smooth. Adjacent development periods typically depend on each other, giving patterns recognisable shapes. However, noise in the data, especially in sparse monthly triangles, or outlying development ratios can produce jagged patterns that the actuary does not regard as credible.

As mentioned in Section~\ref{sec:by-hand}, the typical response is to smooth the estimated development pattern, for example by fitting a parametric curve, applying a regression model, or manually adjusting apparent outliers. The drawback is that none of these approaches measures the goodness of fit given up in exchange for the smoother pattern.

A penalty term makes this trade-off explicit and measurable. The optimisation trades fit to the observed data against smoothness of the fitted development pattern, with the strength of the trade-off governed by a transparent tuning parameter $\alpha_{\text{smooth}}$.

Viewed as a series in development time, the pattern has a natural measure of roughness in the mean squared second difference,
\[
    \frac{1}{n-2} \sum_{j=2}^{n-1} \left(f_{j-1} - 2 f_j + f_{j+1}\right)^2.
\]
This penalty is the standard case of Whittaker-Henderson smoothing \citep{ref:Whittaker1922}.

The penalty inherits the weighting rule of Section~\ref{sec:reference}, for the same reason: carrying the empirical loss's weights $W_j$ keeps the weights from distorting the balance between fit and shape. It drops the normalisation, which the weight $\alpha_{\text{smooth}}$ absorbs:
\begin{align}\label{eq:smooth-penalty}
    J_{\text{smooth}}(\boldsymbol{f}) = \sum_{j=2}^{n-1} W_j \left(2 f_j - f_{j-1} - f_{j+1}\right)^2.
\end{align}

Equation~\eqref{eq:smooth-penalty} changes the structure of the minimisation. The second difference couples each ratio to its neighbours. The objective is therefore no longer separable per development ratio. Appendix~\ref{ap:matrix} builds the matrix formulation, which Section~\ref{sec:objective} then uses as its main result. The appendix writes the weighted loss and the shape penalties as quadratic forms, and derives the normal equations of the combined objective.

Whittaker-Henderson smoothing is long established in actuarial practice \citep{ref:Taylor1992}. The main method in the reserving literature assumes a parametric form for the development ratios. For example, \citet{ref:Sherman1984} smooths development factors by fitting parametric curves, most notably the inverse power curve. \citet{ref:Korn2017} extends this approach through the use of smoothing splines. Although both methods share the objective of reducing noise in development ratios, the mechanism by which smoothness is imposed differs fundamentally from that proposed in this paper. Introducing a penalty term into the objective function optimises smoothness while retaining a non-parametric representation of the development ratios. Consequently, the data are able to override the smoothing where the underlying signal is sufficiently strong.

\begin{figure}[!htbp]
  \centering
  \includegraphics[width=0.75\textwidth]{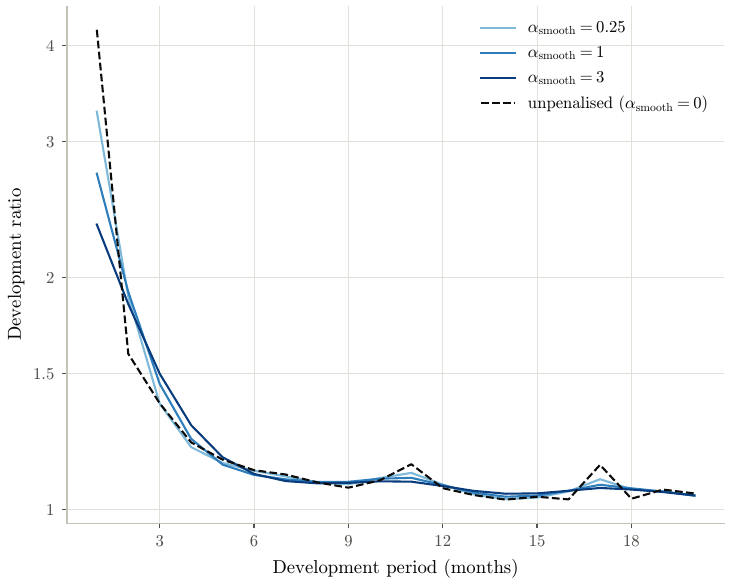}
  \caption{The smoothness penalty on the synthetic illustration triangle. The unpenalised pattern (dashed line) against fitted patterns for increasing $\alpha_{\text{smooth}}$: the roughness is smoothed away while the level and shape of the pattern survive.}
  \label{fig:smoothness}
\end{figure}

Figure~\ref{fig:smoothness} shows the penalty acting on the synthetic illustration triangle, whose generator encodes a rough pattern deliberately. As $\alpha_{\text{smooth}}$ increases, the period-to-period jaggedness is smoothed away while the pattern's level and shape survive.

\FloatBarrier

\subsection{The objective function, assembled}\label{sec:objective}

Sections~\ref{sec:decay}--\ref{sec:smoothness} built the pieces; this section puts them together and solves the result. The core objective is the weighted empirical loss plus the two shape penalties,
\begin{align}\label{eq:assembled-objective}
    \mathscr{L}(\boldsymbol{f})
    = \operatorname{Err}(\boldsymbol{f}; \rho, \gamma)
    + \alpha_{\text{ref}} \, J_{\text{ref}}(\boldsymbol{f})
    + \alpha_{\text{smooth}} \, J_{\text{smooth}}(\boldsymbol{f}),
\end{align}
an instance of the general form of Equation~\eqref{eq:general-form}. Each term is a quadratic in $\boldsymbol{f}$. The empirical term is strictly convex whenever every development period retains positive weight ($W_j > 0$), and both penalties are positive semi-definite. Hence $\mathscr{L}$ is strictly convex and has a unique minimiser. The convexity belongs to the core objective; the shift and rotation penalties of Sections~\ref{sec:shift} and~\ref{sec:rotate} give it up.

In general, an objective of the form of Equation~\eqref{eq:general-form} is minimised numerically. Any penalty that is differentiable almost everywhere is admissible, and gradient-based optimisers handle the rest. This generality is part of the framework. Any adjustment that can be written as such a penalty can enter the fit, and its weight can be trained, whether or not it admits a closed form. For the penalties proposed in Sections~\ref{sec:decay}--\ref{sec:smoothness}, however, no optimiser is needed. The assembled core objective solves in a single linear system, the normal equations derived in Appendix~\ref{ap:matrix},
\begin{align}
    \left[(1 + \alpha_{\text{ref}}) \, \mathrm{W} + \alpha_{\text{smooth}} \, \mathrm{D}^\top \tilde{\mathrm{W}} \mathrm{D}\right] \hat{\boldsymbol{f}}
    = \mathrm{W} \left(\bar{\boldsymbol{f}} + \alpha_{\text{ref}} \, \dot{\boldsymbol{f}}\right).
\end{align}
Each component of the solution is individually interpretable: the reference-pattern term acts as the credibility blend of Equation~\eqref{eq:credibility-blend}, and the smoothness term acts as Whittaker-Henderson smoothing of the blended pattern.

\begin{remark}
Throughout this paper, the solution $\hat{\boldsymbol{f}}$ of the normal equations at given hyperparameters is called the \textup{fitted} pattern. The hyperparameters are inputs to it, not outputs of it; selecting them is a separate exercise, explained in Section~\ref{sec:training}.
\end{remark}


\FloatBarrier

\section{Hyperparameter-based pattern adjustments}\label{sec:adjustments}

The shape or timing of claim developments implied by a development pattern needs adjusting from time to time. The trigger may be an underlying data trend, a business process backlog or change, a regulatory intervention, or a deliberate claims-handling initiative. Left unadjusted, a pattern calculated from historical triangle data becomes unrepresentative of the true underlying process, and the projections built on it run systematically high or low.

Section~\ref{sec:building} assembled the core objective and solved it in closed form. This section extends the framework with three further adjustments. Each addresses the same concern - a change in how quickly claims develop - and each reduces to a single hyperparameter.

The manual counterpart of changing a pattern shape is many simultaneous ratio edits, difficult to document and beyond the reach of an algorithmic selection procedure. Compressed into one declarable parameter, the same adjustment becomes a quantity a reviewer can state and audit. Its application is automatic and repeatable.

Each such hyperparameter can be declared by judgement as an adjustment to the pattern. It can equally be trained by a process like that of Section~\ref{sec:training}, letting the experience data tell the story.

\subsection{The tail-weight shift penalty}\label{sec:shift}

A pattern shape adjustment that adds or reduces weight in the tail can be seen as changing the pattern's implied cash flow duration. For example, it moves the mean settlement time earlier or later. This section writes a tail adjustment as the tail-weight shift penalty.

To express the pattern's development profile we write the percentage developed implied by the ratios as
\[
    \pi_{n+1} = 1, \qquad \pi_j = \frac{\pi_{j+1}}{f_j} = \prod_{m=j}^{n} f_m^{-1},
\]
and the incremental percentage developed - the share of the ultimate amount that emerges in each development period - as
\[
    h_j = \pi_j - \pi_{j-1} = \left(1 - \frac{1}{f_{j-1}}\right) \pi_j \quad \text{for } j > 1,
    \qquad h_1 = \pi_1.
\]
Fitting a simple linear regression of $h_j$ against development time $j$ summarises the development profile's timing in one number. A steep positive slope means the development arrives late, and a shallow or negative slope means it arrives early. The fitted slope is
\[
    \hat\beta(\boldsymbol{f}) = \frac{\sum_{j=1}^{n+1} (j - \bar\jmath)(h_j - \bar h)}{\sum_{j=1}^{n+1} (j - \bar\jmath)^2}.
\]
Two refinements make the slope usable as a penalty. First, negative incremental development would let the optimiser manufacture slope from sign effects. Such development arises from salvage, subrogation and reserve releases in incurred data. The increments are therefore truncated at zero, $h_j^+ = \max(h_j, 0)$, and the slope $\hat\beta^*$ is computed from the truncated increments. Second, the slope is unbounded, so it is passed through $\operatorname{sigmoid}(x) = 1/(1 + e^{-x})$:
\begin{align}\label{eq:shift-penalty}
    J_{\text{shift}}(\boldsymbol{f}) = \operatorname{sigmoid}\left(\hat\beta^*(\boldsymbol{f})\right).
\end{align}
The sigmoid bounds the penalty in $(0,1)$, so the shift term cannot dominate the objective. It does not remove the monotone pull on $\hat\beta^*$, since the penalty always prefers a lower slope. What stops the pattern from shifting indefinitely is the empirical term pulling back. The penalty enters the objective as $\alpha_{\text{shift}} \, J_{\text{shift}}$ with $\alpha_{\text{shift}}$ of either sign: positive to reward earlier development, negative to reward later.

The extension comes at a cost. Since $J_{\text{shift}}$ is not convex in $\boldsymbol{f}$, including it in the objective sacrifices both the convexity guarantee of Section~\ref{sec:objective} and the associated closed-form solution. The resulting objective must therefore be minimised numerically. Nevertheless, the objective remains differentiable almost everywhere. The exception is the truncation operator, which is non-differentiable only when an increment is exactly zero. Appendix~\ref{ap:shift-gradient} derives the complete gradient, so gradient-based optimisation methods can be used to minimise it. More generally, this follows the framework of Section~\ref{sec:objective}: the practitioner specifies a new objective, expresses it as an almost-everywhere differentiable penalty, and optimises it jointly with the existing terms.

\begin{figure}[!htbp]
  \centering
  \includegraphics[width=0.95\textwidth]{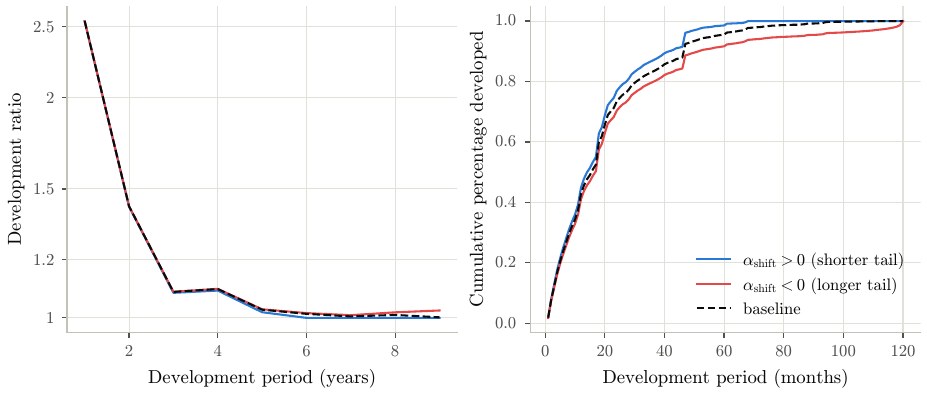}
  \caption{The shift penalty on the synthetic illustration triangle. Left: the fitted development ratios (log scale) on the triangle aggregated to accident years, the baseline (dashed) against the shift-penalised fits. Right: the cumulative percentage developed on the monthly triangle.}
  \label{fig:shift}
\end{figure}

Figure~\ref{fig:shift} shows how the penalty acts on the payout profile of the synthetic illustration triangle. The effect on the development ratios is clearest on the annual triangle. For either sign of $\alpha_{\text{shift}}$, the tail of the pattern gains or loses weight. The percentage developed is shown on the monthly triangle, where the penalty visibly moves the pattern's distinctive jump at development month 45. The empirical term meanwhile constrains the fit to the data and preserves the property that each development pattern converges to one. The penalty acts only on the slope of the incremental development pattern, whereas the empirical term penalises per-period deviations from the data. The least costly adjustment is therefore typically a translation of the profile.

\FloatBarrier

\subsection{The slope rotation penalty}\label{sec:rotate}

Another pattern adjustment done by hand is to steepen or flatten the general slope of the development ratios by eye, until they ``look right''. This is similar to adjusting the ratios so that a line fitted through them rotates. The slope rotation penalty therefore rotates the average slope of the development ratios clockwise or anticlockwise.

Let $g$ be an increasing, almost-everywhere differentiable transformation of a development ratio and $\omega_j > 0$ a weight on development period $j$. Consider the weighted regression of $g(f_j)$ on development time,
\begin{align}\label{eq:rotate-slope}
    \hat\beta_g(\boldsymbol{f}) = \frac{\sum_{j=1}^{n} \omega_j\, (j-\bar\jmath_\omega)\, g(f_j)} {\sum_{j=1}^{n} \omega_j\, (j-\bar\jmath_\omega)^2},
    \qquad
    \bar\jmath_\omega = \frac{\sum_{j=1}^{n} \omega_j\, j}{\sum_{j=1}^{n} \omega_j}.
\end{align}
Passing this slope through the same sigmoid yields
\begin{align}\label{eq:rotate-penalty}
    J_{\text{rot}}(\boldsymbol{f}) = \operatorname{sigmoid}\left(\hat\beta_g(\boldsymbol{f})\right),
\end{align}
which enters the objective as $\alpha_{\text{rot}} \, J_{\text{rot}}$.

This penalty formulation is generic and requires two architectural choices. The first is the weighting $\omega_j$. For comparability the $W_j$ weights are used, multiplied by the index weighting $(n-j+1)^2$, that is, $\omega_j = W_j (n-j+1)^2$. The index weighting is included to concentrate the influence of the rotation on the earlier development ratios. The second is the function $g$. The natural logarithm is used throughout.

\begin{remark}
A good alternative for paid patterns that are strictly positive is $g(x) = \ln\left(\max(x - 1, \epsilon)\right)$, where $\epsilon > 0$ is a small floor. Ratios near one are indistinguishable under $\ln x$, but separated across several orders of magnitude under $\ln(x-1)$. This picks up tail variation.
\end{remark}

We add this term to the objective of Section~\ref{sec:objective}. As in Section~\ref{sec:shift}, the penalty makes the objective non-convex, so it must be minimised numerically. The gradient is, however, straightforward:
\[
    \frac{\partial J_{\text{rot}}}{\partial f_j}
    = \operatorname{sigmoid}\left(\hat\beta_g\right)\left(1 - \operatorname{sigmoid}\left(\hat\beta_g\right)\right)
      \frac{\omega_j\, (j - \bar\jmath_\omega)\, g'(f_j)}{\sum_{m} \omega_m\, (m - \bar\jmath_\omega)^2}.
\]

A positive penalty weight $\alpha_{\text{rot}}$ pushes development ratios in opposite directions on either side of $\bar\jmath_\omega$: ratios before $\bar\jmath_\omega$ increase while those after $\bar\jmath_\omega$ decrease. A negative weight reverses the effect. The pattern therefore pivots around $\bar\jmath_\omega$ rather than translating as a whole. The rotation of the fitted ratio line is clockwise for a positive weight and anticlockwise for a negative one.

\begin{figure}[!htbp]
  \centering
  \includegraphics[width=0.95\textwidth]{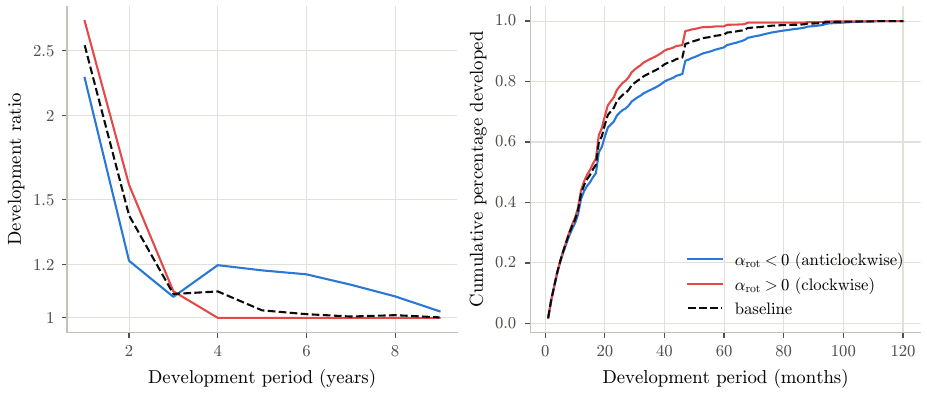}
  \caption{The rotation penalty under the transformation $g(x) = \ln x$ with weights $\omega_j = (n-j+1)^2\, W_j$. Left: the fitted development ratios (log scale) on the same annual triangle as Figure~\ref{fig:shift}. Right: the cumulative percentage developed on the monthly triangle.}
  \label{fig:rotate}
\end{figure}

Figure~\ref{fig:rotate} illustrates rotations in both directions, labelled according to the rotation of the ratio line. The clockwise fit elevates the early ratios, while its tail becomes constrained by the lower bound of one; the anticlockwise fit instead elevates the tail. The right-hand panel applies the same penalty to the monthly triangle and displays the percentage developed, where the impact on the pattern's distinctive jump at month 45 is clearly visible. In this setting, the rotation may be interpreted as a re-timing of the entire development profile.

\FloatBarrier

\subsection{Pattern drift}\label{sec:drift}

Sections~\ref{sec:shift} and~\ref{sec:rotate} introduced techniques to adjust the general shape of the development pattern. When the development pattern changes over time, past data become unrepresentative of today. This is data drift, in machine learning terms. The projection must be adjusted for the future, and the old data become hard to use as they stand. A common example is a change in the claim settlement process over time. In this paper this is referred to as pattern drift.

A pattern fitted to all of the data when there is pattern drift mis-times future development systematically. The most recent diagonals already develop on a different clock compared to what the fitted pattern implies.

This subsection constructs a pattern drift adjustment through a one-parameter rescaling of development time. It can accelerate or decelerate any fitted pattern after the objective has been optimised.

It rescales the pattern by reading the current fitted pattern on a stretched or compressed development clock, mapping period $j$ to a rescaled, or warped, period $\psi_d(j)$. That period is not restricted to being an integer. For drift strength $d$,
\[
    \psi_d(j) = 1 + \frac{n \, e^{d} \, (j-1)}{n + \left(e^{d} - 1\right)(j-1)},
    \qquad j \in [1,\, n+1].
\]
This transformation corresponds to a constant shift of $d$ in the log-odds of development age, expressed as a proportion of the pattern's range. The warp is strictly increasing and fixes both endpoints, $\psi_d(1) = 1$ and $\psi_d(n+1) = n+1$. For $d > 0$ it satisfies $\psi_d(j) \ge j$ throughout, so the pattern is read at older periods than the calendar shows and development is accelerated. For $d < 0$ it is read at younger periods and development is decelerated; $d = 0$ is the identity pattern. Because log-odds shifts compose additively, the family is symmetric about the identity: $\psi_d^{-1} = \psi_{-d}$. Accelerating by $d$ and decelerating by $-d$ are therefore exact inverses of one another, and the drift strength reads on a single scale.

\begin{figure}[!htbp]
  \centering
  \includegraphics[width=0.55\textwidth]{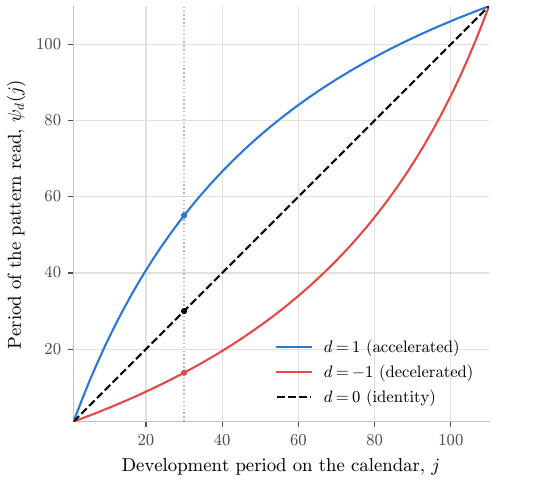}
  \caption{The rescaling $\psi_d$ of development time: which period of the fitted pattern is read at each calendar period, above the identity for $d > 0$ and below it for $d < 0$.}
  \label{fig:warp}
\end{figure}

Figure~\ref{fig:warp} shows the warp itself at $d = 1$, $0$ and $-1$. Its horizontal axis is the development period a ratio belongs to on the calendar. Its vertical axis is the warped period at which the fitted pattern is read. For example, month 30 is read at month 55 as shown in the figure. What the warp bends is therefore the development clock, not the pattern. It is the same fitted pattern throughout, read on a stretched or compressed timescale. Because both endpoints are fixed, the warp stays inside the pattern's own range.

Because the warped periods generally fall between the integer values, reading the fitted pattern there requires it to be a continuous function. Let $F_1 = 1$ and $F_j = \prod_{m=1}^{j-1} f_m$ be the cumulative development factors implied by the fitted ratios. We can extend these factors between the integer periods by geometric interpolation:
\begin{align}\label{eq:drift-interp}
    \widetilde F(u) = F_{\lfloor u \rfloor} \, f_{\lfloor u \rfloor}^{\,u - \lfloor u \rfloor},
\end{align}
so that $\widetilde F$ is continuous, positive and log-linear between periods, with $\widetilde F(n+1) = F_{n+1}$ at the right endpoint. The drifted ratios read the interpolated factors between consecutive warped periods:
\begin{align}\label{eq:drifted-ratios}
    f_j^{*} = \frac{\widetilde F\!\left(\psi_d(j+1)\right)}{\widetilde F\!\left(\psi_d(j)\right)},
    \qquad j = 1, \ldots, n.
\end{align}

Two properties of pattern drift adjustment follow directly. First, at $d = 0$, $f_j^{*} = f_j$: the original fitted pattern is recovered exactly, because time is not warped. Second, drift redistributes development across periods without changing the total development the pattern reaches by its final observed period, because the warp fixes both endpoints:
\[\prod_{j=1}^{n} f_j^{*} = F_{n+1} = \prod_{j=1}^{n} f_j.\]

The parameter $d$ is a natural declarable quantity for a pattern drift adjustment, and it can serve either kind of adjustment of Section~\ref{sec:by-hand}. As a prospective adjustment, $d$ is declared to express a change the triangle cannot yet show. The worked example of Section~\ref{sec:example} declares it this way. As an experience adjustment, $d$ would reflect a re-timing trend already present in the history. Because the warp acts on the fitted pattern and can be scored on the same held-out diagonals as any other hyperparameter, its value could equally be selected by the training described in Section~\ref{sec:training}.

\begin{figure}[!htbp]
  \centering
  \includegraphics[width=0.95\textwidth]{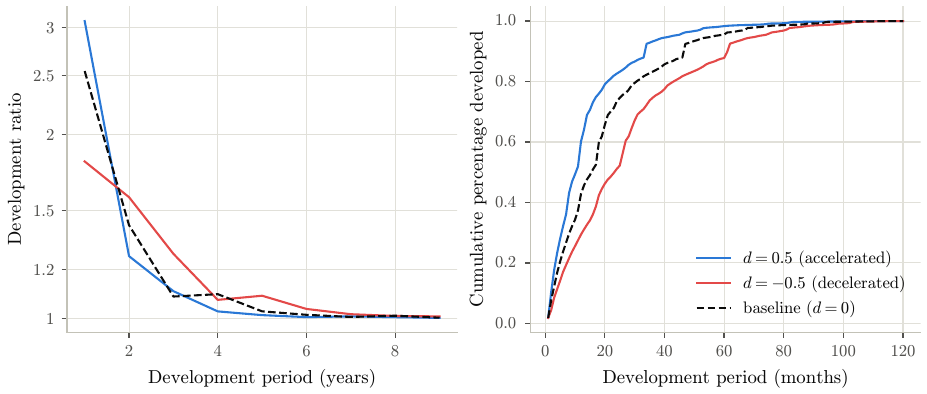}
  \caption{Pattern drift on the synthetic illustration triangle. Left: the fitted development ratios (log scale), the baseline (dashed) against its drifted counterparts. Right: the cumulative percentage developed under the same drifts, on the monthly triangle. $d > 0$ accelerates the development, $d < 0$ decelerates it.}
  \label{fig:drift}
\end{figure}

Figure~\ref{fig:drift} illustrates the mechanism on the synthetic illustration triangle. The left panel plots the yearly development ratio pattern used earlier, to show the effect of accelerating or decelerating development. The right panel shows the pattern's distinctive jump around month 45 and how the warp moves it earlier or later. This contrasts with the penalties earlier in this section, which leave the timing of the pattern's features unchanged.

\section{Training the model}\label{sec:model-training}

An algorithm minimising the objective function in Section~\ref{sec:objective} \textit{fits} the pattern for given hyperparameters. Denote the minimiser as $\hat{\boldsymbol{f}}$. The optimality is restricted to the selected hyperparameters: the loss weights $\rho$ and $\gamma$, the penalty weights $\alpha_{\text{ref}}$ and $\alpha_{\text{smooth}}$, and the drift parameter $d$ of Section~\ref{sec:drift}. These have thus far been specified by judgement.

It is possible instead to select them using a hold-out validation process. This process is referred to as \textit{training} in this paper. Fitting solves for the pattern at given hyperparameters; training selects the hyperparameters themselves.

This section first constructs the validation score that measures held-out prediction. It then develops the training loop that selects every trainable hyperparameter together by that score, the objective's weights and the drift parameter alike. Finally, it demonstrates the loop in an end-to-end example.

\subsection{The reserve validation score}\label{sec:validation-score}

A good training process selects the hyperparameters that best predict future experience. To search for the best candidate, we need a realistic validation score. This subsection constructs that score. Section~\ref{sec:training} applies it.

The first component of the validation score is the split between in-sample and \gls{oos} data. That split should reflect the forecasting task as closely as possible. Because reserving concerns future calendar periods, validation holds out one or more recent calendar diagonals and fits the model to the remaining triangle. The fitted pattern's projections are then scored against the held-out claims, as if the hyperparameters were chosen in an earlier reserving exercise.

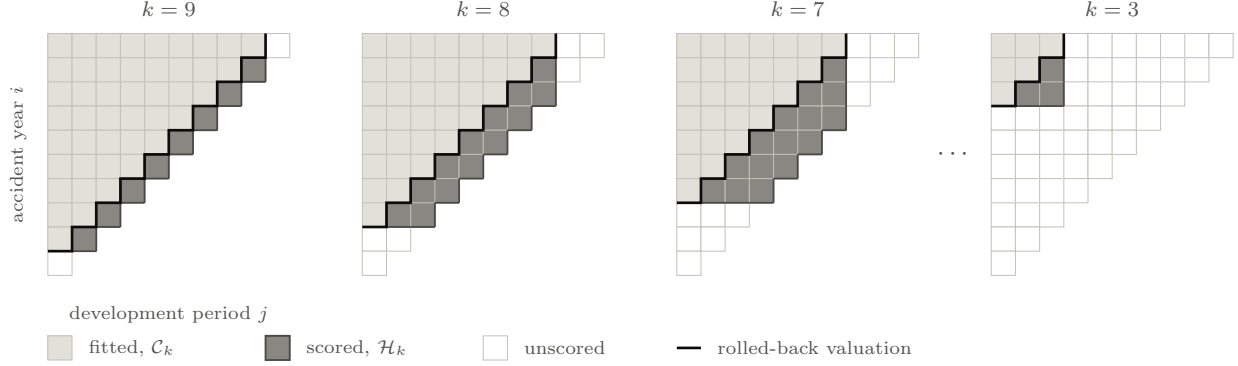
\begin{figure}[!htbp]
  \centering
  \newcommand{\svYears}{10}          
\newcommand{\svPanelList}{9,8,7,3} 
\newcommand{\svFloor}{3}           
\newcommand{\svCellSize}{3.2mm}    
\newcommand{\svGap}{3}             

\newcommand{\svLabelFit}{fitted, $\mathcal{C}_{k}$}
\newcommand{\svLabelScore}{scored, $\mathcal{H}_{k}$}
\newcommand{\svLabelUnused}{unscored}
\newcommand{\svLabelCut}{rolled-back valuation}
\newcommand{\svLabelSplit}[1]{$k = #1$}
\newcommand{\svLabelYear}{accident year $i$}
\newcommand{\svLabelDev}{development period $j$}
\newcommand{\svLabelForcing}{predecessor also ahead of the cut}

\newcommand{\svForceShow}{0}       
\newcommand{\svForcePanel}{8}      
\newcommand{\svForceRow}{3}        
\newcommand{\svForceCol}{7}        

\definecolor{svfit}{HTML}{E1E0D9}
\definecolor{svscore}{HTML}{898781}
\definecolor{svunused}{HTML}{FFFFFF}
\definecolor{svrule}{HTML}{C3C2B7}
\definecolor{svink}{HTML}{0B0B0B}
\definecolor{svmuted}{HTML}{52514E}
\definecolor{svedge}{HTML}{4A4946}
\newcommand{\svEdgeWidth}{0.7pt}


\newcommand{\svDrawCell}[3]{%
  \fill[#3] ({#2-1},{1-#1}) rectangle ({#2},{-#1});%
  \draw[svrule,line width=0.2pt] ({#2-1},{1-#1}) rectangle ({#2},{-#1});%
}

\newcommand{\svScoredInto}[4]{%
  \pgfmathtruncatemacro{#1}{%
    ((#2)>=1 && (#3)>=1 && (#2)+(#3)<=\svYears+1 && (#2)+(#3)-1>(#4)
     && (#2)<=(#4) && (#3)<=(#4)) ? 1 : 0}%
}

\newcommand{\svOutlineCell}[3]{%
  \svScoredInto{\svHere}{#1}{#2}{#3}%
  \ifnum\svHere=1\relax
    \svScoredInto{\svAbove}{#1-1}{#2}{#3}%
    \svScoredInto{\svBelow}{#1+1}{#2}{#3}%
    \svScoredInto{\svLeftN}{#1}{#2-1}{#3}%
    \svScoredInto{\svRightN}{#1}{#2+1}{#3}%
    \ifnum\svAbove=0\relax
      \draw[svedge,line width=\svEdgeWidth] ({#2-1},{1-#1}) -- ({#2},{1-#1});%
    \fi
    \ifnum\svBelow=0\relax
      \draw[svedge,line width=\svEdgeWidth] ({#2-1},{-#1}) -- ({#2},{-#1});%
    \fi
    \ifnum\svLeftN=0\relax
      \draw[svedge,line width=\svEdgeWidth] ({#2-1},{1-#1}) -- ({#2-1},{-#1});%
    \fi
    \ifnum\svRightN=0\relax
      \draw[svedge,line width=\svEdgeWidth] ({#2},{1-#1}) -- ({#2},{-#1});%
    \fi
  \fi
}

\newcommand{\svCell}[3]{%
  \pgfmathtruncatemacro{\svI}{#1}%
  \pgfmathtruncatemacro{\svJ}{#2}%
  \pgfmathtruncatemacro{\svK}{#3}%
  \pgfmathtruncatemacro{\svD}{\svI+\svJ-1}%
  \pgfmathtruncatemacro{\svN}{\svYears}%
  \ifnum\svD>\svN\relax
  \else
    \ifnum\svD>\svK\relax
      \ifnum\svI>\svK\relax
        \svDrawCell{\svI}{\svJ}{svunused}
      \else
        \ifnum\svJ>\svK\relax
          \svDrawCell{\svI}{\svJ}{svunused}
        \else
          \svDrawCell{\svI}{\svJ}{svscore}%
        \fi
      \fi
    \else
      \svDrawCell{\svI}{\svJ}{svfit}%
    \fi
  \fi
}

\newcommand{\svPanel}[2]{%
  \begin{scope}[shift={(#2,0)}]
    \pgfmathtruncatemacro{\svCut}{#1}%
    \foreach \svRow in {1,...,\svYears}{%
      \foreach \svCol in {1,...,\svYears}{%
        \svCell{\svRow}{\svCol}{\svCut}%
      }%
    }%
    \foreach \svRow in {1,...,\svYears}{%
      \foreach \svCol in {1,...,\svYears}{%
        \svOutlineCell{\svRow}{\svCol}{\svCut}%
      }%
    }%
    \draw[svink,line width=1.0pt,line join=miter]
      (\svCut,0) \foreach \svStep in {1,...,\svCut}{%
        -- ({\svCut-\svStep+1},{-\svStep}) -- ({\svCut-\svStep},{-\svStep})};
    \node[svmuted,font=\footnotesize,anchor=south]
      at ({\svYears/2},{0.35}) {\svLabelSplit{#1}};
  \end{scope}%
}


\begin{tikzpicture}[x=\svCellSize,y=\svCellSize,>={Stealth[length=1.4mm]}]

  \pgfmathtruncatemacro{\svStride}{\svYears+\svGap}%
  \pgfmathsetmacro{\svFootY}{-\svYears-0.9}%
  \pgfmathsetmacro{\svAnnY}{-\svYears-2.1}%

  \foreach \svS [count=\svIdx from 0,
                 remember=\svS as \svPrev (initially \svYears)] in \svPanelList {%
    \svPanel{\svS}{\svIdx*\svStride}%
    \ifnum\svIdx>0\relax
      \pgfmathtruncatemacro{\svJump}{\svPrev-\svS}%
      \ifnum\svJump>1\relax
        \node[svmuted] at ({\svIdx*\svStride-\svGap/2},{-\svYears/2}) {$\cdots$};
      \fi
    \fi
    \ifnum\svIdx=0\relax
      \node[svmuted,font=\scriptsize,rotate=90,anchor=south]
        at ({-0.5},{-\svYears/2}) {\svLabelYear};
      \node[svmuted,font=\scriptsize,anchor=north]
        at ({\svYears/2},{\svFootY}) {\svLabelDev};
    \fi
    \ifnum\svForceShow=1\relax
      \ifnum\svS=\svForcePanel\relax
        \begin{scope}[shift={(\svIdx*\svStride,0)}]
          \draw[svink,line width=0.6pt]
            ({\svForceCol-1},{1-\svForceRow})
            rectangle ({\svForceCol+1},{-\svForceRow});
          \draw[svink,line width=0.6pt,->]
            ({\svForceCol-0.5},{0.5-\svForceRow})
            -- ({\svForceCol+0.5},{0.5-\svForceRow});
          \draw[svmuted,line width=0.4pt,dotted]
            ({\svForceCol+1},{-\svForceRow}) -- ({\svForceCol+1},{\svAnnY+0.35});
          \node[svmuted,font=\scriptsize,anchor=north]
            at ({\svForceCol+1},{\svAnnY+0.35}) {\svLabelForcing};
        \end{scope}
      \fi
    \fi
  }%

  \ifnum\svForceShow=1\relax
    \pgfmathsetmacro{\svLegY}{-\svYears-3.9}%
  \else
    \pgfmathsetmacro{\svLegY}{-\svYears-2.5}%
  \fi
  \fill[svfit] ({0},{\svLegY}) rectangle ++(1,-1);
  \draw[svrule,line width=0.2pt] ({0},{\svLegY}) rectangle ++(1,-1);
  \node[svmuted,font=\scriptsize,anchor=west] at ({1.3},{\svLegY-0.5}) {\svLabelFit};
  \fill[svscore] ({9},{\svLegY}) rectangle ++(1,-1);
  \draw[svedge,line width=\svEdgeWidth] ({9},{\svLegY}) rectangle ++(1,-1);
  \node[svmuted,font=\scriptsize,anchor=west] at ({10.3},{\svLegY-0.5}) {\svLabelScore};
  \fill[svunused] ({18},{\svLegY}) rectangle ++(1,-1);
  \draw[svrule,line width=0.2pt] ({18},{\svLegY}) rectangle ++(1,-1);
  \node[svmuted,font=\scriptsize,anchor=west] at ({19.3},{\svLegY-0.5}) {\svLabelUnused};
  \draw[svink,line width=1.0pt] ({26},{\svLegY-0.5}) -- ++(1,0);
  \node[svmuted,font=\scriptsize,anchor=west] at ({27.3},{\svLegY-0.5}) {\svLabelCut};

\end{tikzpicture}
  \caption{The validation splits on a ten-year triangle, for $k = I - 1$ down to $3$. Split $k$ is fitted to the triangle as it stood at calendar period $k$ ($\mathcal{C}_k$), and scored on the transitions that triangle can project forward ($\mathcal{H}_k$).}
  \label{fig:validation-splits}
\end{figure}

Figure~\ref{fig:validation-splits} illustrates the hold-out framework this paper uses. It is the rolling procedure of \citet{ref:CDRScore}, who re-reserve at each past calendar period and score the result on the next diagonal. We make one validation split per historic calendar period, from $I-1$ down to three, the earliest calendar period considered. Each split predicts its valid \gls{oos} observations, and those predictions score the method. The scores from the individual splits are then aggregated.

What makes a hold-out observation valid? The chain ladder predictive framework requires at least one observation for each accident period, which disqualifies many of the newest accident periods. The framework also requires at least one observation in each development period to estimate its ratio. That limits the hold-out data in the other dimension too. So older calendar periods work with much smaller datasets to obtain their scores. For split $k$, let $\mathcal{C}_k$ be the indices of the in-sample data used for model fitting and $\mathcal{H}_k$ the \gls{oos} data. Let $\hat f_j^{(k)}$ be the development ratios fitted on $\mathcal{C}_k$.

The next question is which score to use. It should prioritise the right observations and not be swayed by noise. Many metrics for prediction error on the held-out claims would serve. \citet{ref:CDRScore} score each re-reserving exercise on the next diagonal. Their scores are the actual-versus-expected increments and the claims development result, each averaged over accident years with claim-size weights. The second also penalises the change in the reserve between valuations. This paper scores each held-out development transition as a one-step standardised residual, under the variance assumption of \citet{ref:Mack1993}. Dividing by $C_{i,j}\sigma_j^2$ reduces the influence of noisy claims.

The variance parameters are estimated by Mack's estimator, applied to the unpenalised \gls{vwcl} residuals of the full triangle. Every raw estimate is then replaced by a smoothed one. We fit a straight line to $\log \hat\sigma_j$ in $j$, weighting each development period by its degrees of freedom. We then read that line at every period. \citet{ref:Mack1993} obtains the final $\sigma_j$ by log-linear extrapolation, because the estimates usually decay exponentially in $j$. Smoothing extends that reasoning from the last period to the whole series. It matters most near the tail, where a period carries one or two observations. The fitted line also supplies a value for the last development period, which has no estimate of its own.

The score per split used in this paper is defined as:
\begin{align}\label{eq:val-score}
    \text{score}_k \;=\; \frac{1}{\lvert \mathcal{H}_k \rvert} \sum_{(i,j) \in \mathcal{H}_k} \frac{\bigl( C_{i,j+1} - C_{i,j}\, \hat f_j^{(k)} \bigr)^2}{\sigma_j^2\, C_{i,j}}.
\end{align}
The score is built to isolate each development ratio's contribution to the error. Every transition is therefore projected from the true $C_{i,j}$, even where that claim lies ahead of split $k$'s valuation. That scores each transition on its own, so an error in one diagonal cannot compound into the diagonals after it - a mechanism similar to teacher forcing \citep{ref:WilliamsZipser1989}. The choice does leak the held-out predecessor into the score. The leak reaches only the validation score, never the fit. Every candidate is fitted on the reduced triangle alone and scored on the same fixed weights, so no candidate gains an advantage the others do not share.

To use as much information as possible, we aggregate the scores of all the splits. \citet{ref:CDRScore} average them over the calendar periods. Here the scores are instead weighted by a decay factor $\rho_{\text{score}}^{\,I-k-1}$ to prioritise newer data, yielding
\begin{align}\label{eq:val-combined}
    \text{score} \;=\; \frac{\sum_{k=3}^{I-1} \rho_{\text{score}}^{I-k-1}\; \text{score}_k}{\sum_{k=3}^{I-1} \rho_{\text{score}}^{I-k-1}},
\end{align}
where $\text{score}_k$ is Equation~\eqref{eq:val-score} evaluated on split $k$.

\subsection{The training loop}\label{sec:training}

Section~\ref{sec:objective} expressed actuarial judgement through a set of declarable hyperparameters. Although these may be chosen directly by the actuary, they can also be selected empirically by a supervised-learning training and validation loop \citep{ref:elementsSLT, ref:CDRScore}. This subsection demonstrates such a loop, with the validation score of Section~\ref{sec:validation-score} as its measure.

Training in this sense does not improve the in-sample fit. The penalties in the objective function worsen it by construction, because they pull the pattern away from the weighted empirical solution (Section~\ref{sec:by-hand}). What training improves is prediction on held-out data. The trained weights are the ones whose adjustments would have predicted the triangle's own later diagonals best. The search space comprises the loss weights $\rho$ and $\gamma$, the smoothness weight $\alpha_{\text{smooth}}$, the reference weight $\alpha_{\text{ref}}$, and any other hyperparameter whose predictive value can be evaluated.

Prospective-adjustment weights, however, lie outside this loop. Because prospective adjustments encode views of future experience that are not represented in the observed triangle, validation data cannot inform their calibration. They are therefore \textit{declared} rather than trained - fixed by the practitioner's judgement, without validation.

\begin{remark}
Being trainable does not oblige a hyperparameter to be trained. Any of them may still be fixed a priori, as in Section~\ref{sec:example}, where $\gamma = 1$ is declared throughout.
\end{remark}

Hyperparameter selection can be performed using a training loop. Define a grid of possible hyperparameters. Fit the model at every combination, then select the combination that minimises the \gls{oos} score. Typically the grid is defined on a logarithmic scale because penalty weights' influence tends to scale exponentially.

The loop trains every hyperparameter together, on one grid. A candidate is a point $(\rho, \gamma, \alpha_{\text{ref}}, \alpha_{\text{smooth}})$ of that grid. Algorithm~\ref{alg:training-loop} states the loop in full. Once all the combinations have their validation score, the hyperparameters corresponding to the minimum are selected for the final trained model \citep{ref:elementsSLT}.
\begin{algorithm}[!htbp]
\caption{The training loop}\label{alg:training-loop}
\begin{algorithmic}[1]
\Require triangle $\mathcal{C}$ of $I$ accident years; reference pattern $\dot{\boldsymbol{f}}$; score decay $\rho_{\text{score}}$; hyperparameter grid $G$ over $(\rho, \gamma, \alpha_{\text{ref}}, \alpha_{\text{smooth}})$
\State $\sigma_j \gets$ the smoothed full-triangle estimates of Section~\ref{sec:validation-score}
\State $\mathcal{C}_k \gets \mathcal{C}$ as it stood at calendar period $k$, $\mathcal{H}_k \gets$ the transitions ahead of it, for each $k = I-1, \dots, 3$ \Comment{candidate-independent, prepared once}
\For{each candidate $w \in G$}
    \State $\hat{\boldsymbol{f}}^{(k)} \gets$ the pattern fitted to $\mathcal{C}_k$ using hyperparameters $w$, for each $k$ \Comment{Equation~\eqref{eq:normal-equations}}
    \State $\text{score}(w) \gets$ the $\rho_{\text{score}}$-combined score over $k$ using $\hat{\boldsymbol{f}}^{(k)}$ \Comment{Equations~\eqref{eq:val-score} and~\eqref{eq:val-combined}}
\EndFor
\State $w^{*} \gets \arg\min_{w \in G} \text{score}(w)$
\State \Return $w^{*}$, and the pattern refitted to $\mathcal{C}$ at $w^{*}$
\end{algorithmic}
\end{algorithm}

Evaluating and choosing reserving methods by backtesting on held-out parts of the triangle is common in practice and well developed in literature \citep{ref:Skurnick1973, ref:JingLebensLowe2009, ref:GabrielliWuthrich2018, ref:McGuireTaylorMiller2021}. \citet{ref:CDRScore} took this a step further with a framework that scores a range of candidate reserving models and adjustments on held-out diagonals and selects the best. This paper adapts that framework in two ways. The candidates are not different reserving methods but points on a smooth hyperparameter surface of one chain ladder model's objective function. Each candidate is scored by the one-step standardised residual of Section~\ref{sec:validation-score} rather than by an actual-versus-expected or claims development result score.

Figure~\ref{fig:validation-surface} shows the validation-error surface for the incurred stage of the worked example in Section~\ref{sec:example}. It summarises held-out \gls{oos} error as a function of the hyperparameters. The figure is a two-dimensional slice through the trained hyperparameter space, taken at the selected values of the remaining hyperparameters. For this triangle, the minimiser lies in the interior of the declared grid on both axes.

The resulting surface shows the fit-versus-objective trade-off of Section~\ref{sec:by-hand}. Error rises both towards the unpenalised fit and towards over-penalisation, along the reference and smoothness axes alike.

\FloatBarrier

\subsection{A worked example: fitting with supervised constraints}\label{sec:example}

\begin{figure}[H]
  \centering
  \includegraphics[width=0.95\textwidth]{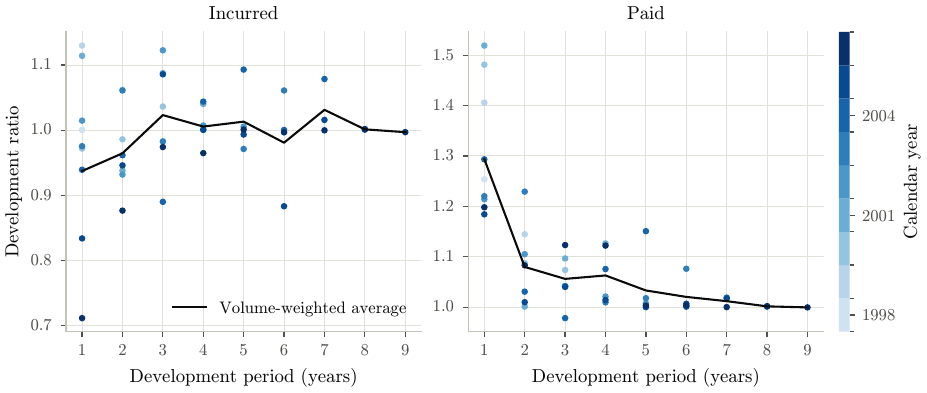}
  \caption{The worked dataset's individual incurred (left) and paid (right) development ratios, coloured by calendar year, shaded oldest (light) to newest (dark) blue, with each triangle's volume-weighted average overlaid.}
  \label{fig:development-ratios}
\end{figure}

This section takes one Schedule P dataset through a two-stage reserving pattern selection, as a demonstration of training with supervised constraints. The class is Other Liability from Tennessee Farmers Mutual, group code 6947, a small insurer whose claim volumes leave the class's own experience noisy. In this setting it is common for an actuary to reach for industry credibility and smoothing rather than trust the \gls{vwcl} pattern alone. The dataset, shown in Figure~\ref{fig:development-ratios}, was hand-picked to display what this paper's supervised-learning framework is built to address.

The unpenalised \gls{vwcl} pattern is visibly rough, and the individual development ratios $f_{i,j}$ are noisy. This is a typical scenario for experience adjustments: excluding outliers, smoothing the pattern, and shaping it towards a more credible industry benchmark. On visual inspection the data carry no obvious trends beyond incurred claim estimates declining and payment rates slowing down.

Two inputs to the training process are declared up front. The validation score's recency decay is $\rho_{\text{score}} = 0.90$, so the \gls{oos} scores of deeper roll-back splits count less than recent ones. For familiarity, the volume power is held at $\gamma = 1$, keeping the \gls{vwcl}'s volume weighting throughout.

\begin{figure}[H]
  \centering
  \includegraphics[width=0.75\textwidth]{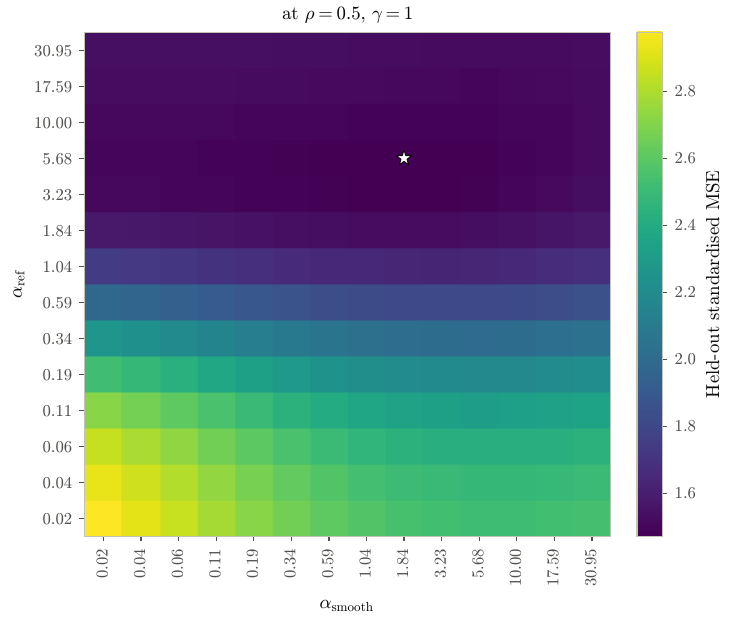}
  \caption{The incurred stage's validation-error surface: \gls{oos} error over the $(\alpha_{\text{ref}}, \alpha_{\text{smooth}})$ plane, with $\gamma$ at its declared value. Error rises towards the unpenalised pattern on one side of the minimiser and towards over-penalisation on the other.}
  \label{fig:validation-surface}
\end{figure}

The incurred chain ladder is trained first, by the loop of Section~\ref{sec:training}. The calendar-period trends just noted call for the recency weighting of Section~\ref{sec:decay}, and the rough pattern for the smoothness penalty of Section~\ref{sec:smoothness}. Because the class's own experience is thin, an industry pattern calibrated on other Schedule P data serves as the reference of Section~\ref{sec:reference}. Training selects $\rho = 0.50$, $\alpha_{\text{ref}} = 5.68$ and $\alpha_{\text{smooth}} = 1.84$. Figure~\ref{fig:validation-surface} maps the held-out score over the $\alpha$ plane at that $\rho$, with the minimiser marked in the grid's interior. The map is a two-dimensional slice of the joint validation-error surface of Section~\ref{sec:training}.

The trained weights read as statements about the data and the projection process. The material $\alpha_{\text{ref}}$ says the industry experience is informative for this company. The optimum in $\alpha_{\text{smooth}}$ says some period-to-period roughness is noise to smooth away.

\begin{figure}[H]
  \centering
  \includegraphics[width=0.75\textwidth]{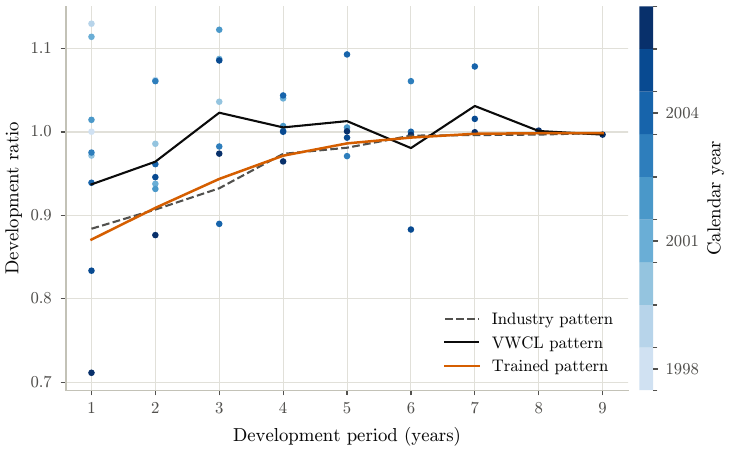}
  \caption{The incurred triangle's individual development ratios (blue), with the unpenalised \gls{vwcl} pattern, the trained pattern ($\rho = 0.50$, $\alpha_{\text{ref}} = 5.68$, $\alpha_{\text{smooth}} = 1.84$), and the declared industry reference (dashed).}
  \label{fig:incurred-patterns}
\end{figure}

These hyperparameters state the experience adjustments on an auditable, objective basis, in place of the manual pattern edits of Section~\ref{sec:by-hand}. Figure~\ref{fig:incurred-patterns} overlays the pattern fitted at the trained weights on the individual incurred ratios. It pulls the jagged \gls{vwcl} pattern smoothly towards the industry pattern.

For further illustration, the trained incurred pattern is used to construct a companion reference for the paid projection, following Section~\ref{sec:reference}. Writing $q_j$ for the company's own paid-to-incurred column ratio at development period $j$, over the accident periods observed at both $j$ and $j+1$, and $\hat f^{\,I}_j$ for the selected incurred pattern,
\[
    \dot f_j \;=\; \frac{q_{j+1}}{q_j}\,\hat f^{\,I}_j \;=\; \bar f^{\,P}_j \, \frac{\hat f^{\,I}_j}{\bar f^{\,I}_j},
\]
with $\bar f^{\,P}_j$ and $\bar f^{\,I}_j$ the paid and incurred triangles' own volume-weighted developments. The reference therefore carries the paid triangle's own development ratios, scaled by exactly the adjustments the incurred training applied. What transfers between the two projections is calibrated judgement.

Substituting the raw incurred \gls{vwcl} pattern $\bar f^{\,I}_j$ for the fitted $\hat f^{\,I}_j$ collapses the reference to the paid \gls{vwcl} pattern $\bar f^{\,P}_j$, which carries no companion information at all.

To train the model on paid data, the same loop as in the incurred stage is run, with this companion reference in place of the industry pattern. Training selects $\rho = 0.15$, $\alpha_{\text{ref}} = 0.59$ and $\alpha_{\text{smooth}} = 10.0$, all interior to the declared grids.

The heavy decay leans on the recent accident years. Their early paid ratios sit visibly below the oldest years' in Figure~\ref{fig:development-ratios}. The strong smoothing irons out the roughness the small class carries.

\begin{figure}[H]
  \centering
  \includegraphics[width=0.75\textwidth]{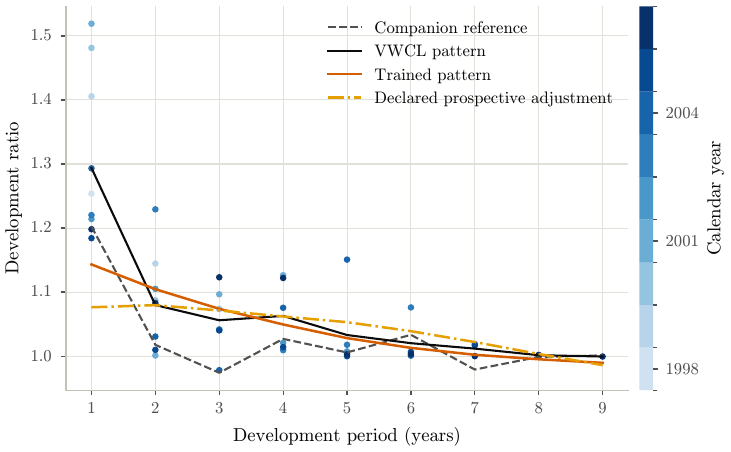}
  \caption{The paid triangle's individual development ratios (blue), with the unpenalised \gls{vwcl} pattern, the trained pattern ($\rho = 0.15$, $\alpha_{\text{ref}} = 0.59$, $\alpha_{\text{smooth}} = 10.0$), the raw companion reference pattern (dashed), and the declared prospective adjustment, the trained pattern re-timed by $d = -0.65$ (dash-dotted).}
  \label{fig:paid-patterns}
\end{figure}

Figure~\ref{fig:paid-patterns} overlays the trained paid pattern on the individual paid ratios. Also shown are the unpenalised \gls{vwcl} pattern and the companion reference derived from the incurred projection. The trained pattern's low first ratio follows the recent years' slower early payments. The \gls{vwcl} pattern averages these together with those of the older years.

The actuary can inspect and override any of these hyperparameters. For instance, a larger $\rho$ can be declared where the decay is thought to be too steep.

Suppose the actuary holds the view that settlement will lengthen by about four months - a claims-handling change, a court backlog - for reasons beyond the slowdown the triangle already shows. Section~\ref{sec:by-hand} classifies such a view as a prospective adjustment. The held-out diagonals are themselves experience, so no training loop can test the view. Once the paid pattern has been trained, the actuary therefore makes this hypothetical judgement call through the pattern drift of Section~\ref{sec:drift}.

The drift strength is solved so that the trained pattern's mean settlement time lengthens by four months, giving $d = -0.65$, a deceleration. The re-timed pattern is illustrated in Figure~\ref{fig:paid-patterns}. It reads every development transition at a younger development age, so less of the ultimate is paid early and more is left to the later periods. The total development the pattern reaches is unchanged.

Finally, the projection options are compared. Table~\ref{tab:reserves} gives the figure an actuary would book under each option, every reserve stated on one paid basis against the run-off that actually emerged. Both \gls{vwcl} patterns overshoot the reserve that emerged. Both trained projections land closer, and the trained paid projection lands closest. As expected, the hypothetical prospective judgement call results in a higher reserve estimate.

\begin{table}[H]
    \centering
    \begin{tabular}{lrrr}
        \toprule
        \rowcolor{gray!15} Projection & Total ultimate & Reserve & Reserve error \\
        \midrule
        \gls{vwcl} incurred                         & 41,541 & 5,458 & $+87.2\%$ \\
        Trained incurred pattern                    & 38,083 & 2,000 & $-31.4\%$ \\
        \gls{vwcl} paid                             & 40,374 & 4,291 & $+47.2\%$ \\
        Trained paid pattern                        & 38,967 & 2,884 & $-1.1\%$ \\
        Declared prospective adjustment             & 40,621 & 4,538 & $+55.7\%$ \\
        \midrule
        Actual run-off                              & 38,998 & 2,915 & n/a \\
        \bottomrule
    \end{tabular}
    \caption{Booked outcomes for each projection option, in thousands of dollars. Every reserve is the projected total ultimate minus the cumulative paid at the valuation date (36,083), and the reserve error is the booked reserve divided by the actual remaining run-off, minus one.}
    \label{tab:reserves}
\end{table}

What this example shows is the mechanics of fitting with supervised constraints on a dataset selected to display them, not a validated claim about Other Liability reserving in general. Whether trained constraints improve on manual selection at portfolio scale (across the full set of Schedule P company-line datasets) remains future work, as set out in Section~\ref{sec:discussion}.

\FloatBarrier

\section{Discussion and limitations}\label{sec:discussion}

\subsection{Penalty interactions}

Declared objectives may conflict. The clearest example is the interaction between the reference-pattern term and a penalty that shifts or alters the shape of the pattern. If the declared reference pattern implies one characteristic while the declared shift penalty favours another, the two terms pull the fitted solution in opposite directions. Another example is the reference pattern conflicting with the smoothness penalty, when the benchmark itself is rougher than the smoothness constraint permits.

This tension between penalties can also be viewed as a diagnostic, not only a modelling nuisance. Nonetheless, the tension can be undesirable. It imposes conflicting readings of the data, and it increases the effect of the penalties, which reduces the relative weight of the empirical term in the objective function.

\subsection{The bootstrap}

The bootstrap is a cornerstone of reserve uncertainty estimation and is widely applied in actuarial reserving \citep{ref:EnglandVerrall1999,ref:ODPBoot}. A robust application relies on the plug-in principle \citep{ref:EfronBootstrap}: the estimated parameters are direct functions of the observed data. Manual adjustments to development ratios violate that principle. The methodology proposed in this paper encodes subjective adjustments in a form that preserves the plug-in principle in the estimation of development ratios. A constrained objective function for the development ratios can therefore let the bootstrap reflect both the underlying data and the actuary's subjective view more faithfully. Applying this approach to the bootstrap is beyond the scope of this paper and is a topic for future research.

\subsection{The scope of the evidence presented}

The mechanisms of Sections~\ref{sec:decay}--\ref{sec:smoothness} and the adjustments of Section~\ref{sec:adjustments} are each illustrated on the synthetic illustration triangle of Section~\ref{sec:running-triangle}. Its data-generating process deliberately encodes the defects those mechanisms are intended to address. For example, a rough pattern is simulated to demonstrate the smoothness penalty.

The worked example of Section~\ref{sec:example}, by contrast, uses a single real Schedule P triangle. The example was selected because its class is small and noisy, its unpenalised pattern is visibly rough, and its paid experience departs from the industry paid pattern, which gives the companion reference mechanism something to demonstrate. Out-of-sample reserve performance was not used to select the class or to train the model. The trained paid pattern landed spuriously close to the actual \gls{oos} reserve, and that closeness should not be read as a sign of the training approach's strength or robustness.

An empirical study scoring the proposed pattern training against manual pattern selection is beyond the scope of this paper.

\subsection{Trained versus declared weights}

The training loop of Section~\ref{sec:training} records only what the data contain, and the two kinds of adjustment of Section~\ref{sec:by-hand} sit on opposite sides of that line. An experience adjustment claims that the triangle misrepresents its own generating mechanism, and the claim is testable. A prospective adjustment asserts that the future will differ from the experience data. Held-out diagonals in the training loop are themselves experience data. A hyperparameter weight trained to predict them is rewarded for reading the past better, not for anticipating a change the past does not contain. Training certain penalty weights can detect trends in the past data, but care should be taken not to train adjustments that are truly prospective.

Declaring a hyperparameter after training expresses the practitioner's subjective view of the future, and is therefore a prospective adjustment. Declaring before training instead encourages the model to learn a more stable representation of the data and prevents overfitting. Declaration before training is therefore an experience adjustment, not a prospective one.

\subsection{When the supervised-learning framework fails}

The framework in this paper replaces manual edits that encode beliefs about the chain ladder's \textit{pattern}. It does not attempt to replace or improve the projection-to-ultimate mechanism. The limitations of the chain ladder therefore still apply.

Three situations sit outside the reach of the training framework of Section~\ref{sec:training}.
The first is small triangles. With few diagonals, holding out enough claims for a stable validation score leaves too little data to train on, which increases the risk of overfitting.
The second is genuinely unstable books. Where the process changes faster than the triangle accrues, no searched objective function survives from one held-out diagonal to the next. In these two scenarios the training process is not robust, and the practitioner can instead declare the hyperparameters by judgement.

The third is errors or misleading information in the triangle. The smoothness penalty should help here compared with the \gls{vwcl}, but misleading or incorrect data can still lead to overfitting and poor generalisation.

Companion data, as described in Section~\ref{sec:reference}, can also reduce fitting to errors in the data. The chain ladder is still likely to require correcting the triangle so that it projects from an accurate view of the latest diagonal, for example by using the paid-incurred gap or the techniques described by \citet{ref:BerquistSherman1977}. The Munich chain ladder \citep{ref:QuargMack2004}, and models that fit the paid and outstanding views jointly \citep{ref:Morris2016, ref:GesmannMorris2020}, improve the projection-to-ultimate mechanism itself, making it more robust to errors in the triangle data.

\section{Conclusion}\label{sec:conclusion}

This paper argues that the adjustments actuaries make to chain ladder patterns by hand are already a form of model training. Each adjustment moves the fitted pattern towards the actuary's view of the mechanism that generates future claim developments. That view is an objective the loss function does not contain. Section~\ref{sec:fitted-model} gave the basis for that reading. The \gls{vwcl} pattern minimises an explicit volume-weighted loss function, so the chain ladder is a fitted model and pattern selection can be treated as a supervised-learning problem. Section~\ref{sec:by-hand} set out and formalised the manual adjustments, showing that a factor selected by judgement is an implicit model choice. It mapped every adjustment onto two mechanisms: reweighting the data inside the empirical loss, and penalising the shape of the pattern through terms added to it.

Section~\ref{sec:building} builds the objective term by term, each derived from the typical manual practice it replaces. Decay and power parameters generalise the data weights for recency and volume (Section~\ref{sec:decay}). The reference-pattern penalty acts as a credibility blend of the weighted empirical solution and the reference pattern (Section~\ref{sec:reference}). The smoothness penalty carries Whittaker-Henderson smoothing of the pattern into the objective (Section~\ref{sec:smoothness}). Assembled, the core objective is strictly convex and solves in a single linear system (Section~\ref{sec:objective}).

The objective function framework extends past the closed form. The tail-weight shift and slope rotation penalties (Sections~\ref{sec:shift} and~\ref{sec:rotate}) reshape the pattern at the cost of convexity and a closed-form solution. The objective remains differentiable almost everywhere, so gradient-based optimisation can still be used to fit it. Pattern drift acts after the development ratios are fitted on the objective, re-timing the fitted pattern (Section~\ref{sec:drift}). Its single parameter accelerates or decelerates future development.

Declaring interpretable weights lets the actuary apply judgement. The same hyperparameters can also be selected objectively from historical data. Holding out calendar diagonals turns hyperparameter selection into a familiar supervised-learning training loop (Section~\ref{sec:model-training}). Every candidate is scored by \gls{oos} error, and the trainable hyperparameters are selected together. The worked example ran the loop end to end on a Schedule P dataset (Section~\ref{sec:example}). It trained the incurred pattern, carried it into the paid projection as a companion reference (Section~\ref{sec:companion}), and scored both against the actual run-off. Held-out diagonals are themselves experience, so experience adjustments can be trained, while prospective adjustments must stay declared.

The practical consequence is a relocation of judgement, not its removal. The actuary still weights recent diagonals over older ones, still distrusts a jagged pattern, and still leans on a benchmark where the book is thin. Those views were previously expressed as edits to individual ratios, through small steps that are hard to report and govern. They are now the declared and trained weights of one objective function, as recorded in Section~\ref{sec:example}. The reserve review and audit conversation changes to comparing a handful of declared weights and the objectives they represent. Where the triangle carries enough data for the training loop, the weights are validated rather than debated.

The framework has limits, set out in Section~\ref{sec:discussion}. Small triangles leave too little data for stable validation. Books that change faster than the triangle accrues outrun any searched objective function. Information that is not a function of the triangle cannot be trained and must be declared by judgement.

Within those limits, the framework is open. Any adjustment that can be written as an almost-everywhere differentiable function of the pattern enters the objective on the same terms as the penalties built here. Companion data enter through a reference pattern derived from them (Section~\ref{sec:reference}). Three refinements are left for future work. Volume-varying penalty weights would let the reference pattern bind hardest where the data are thin. A portfolio-scale evaluation would compare the trained patterns against patterns selected by hand across many real triangles. Carrying the declared objective into the bootstrap would let reserve uncertainty reflect the actuary's subjective view alongside the data.

\clearpage
\printnoidxglossary[type=\acronymtype,title={List of acronyms}]

\bibliography{references}
\newpage
\appendix

\section{Matrix construction and closed-form solutions}\label{ap:matrix}

This appendix rewrites the weighted empirical loss and the shape penalties as quadratic forms and derives the normal equations whose solution is the fitted pattern of Section~\ref{sec:objective}. Throughout, $w_{i,j} = \rho^{\,I-i-j} C_{i,j}^{\gamma}$, $W_j = \sum_{i=1}^{I-j} w_{i,j}$, and
\[
    \bar f_j = \frac{\sum_{i=1}^{I-j} w_{i,j} \, f_{i,j}}{W_j}
\]
is the weighted empirical solution of Equation~\eqref{eq:weighted-solution}.

\subsection{The empirical loss as a quadratic form}

Fix a development period $j$ and expand the inner sum of the weighted loss:
\begin{align}
    \sum_{i=1}^{I-j} w_{i,j} \left(f_{i,j} - f_j\right)^2
    &= \sum_{i=1}^{I-j} w_{i,j} f_{i,j}^2 + f_j^2 \sum_{i=1}^{I-j} w_{i,j} - 2 f_j \sum_{i=1}^{I-j} w_{i,j} f_{i,j} \notag \\
    &= \sum_{i=1}^{I-j} w_{i,j} f_{i,j}^2 + f_j^2 \, W_j - 2 f_j \, W_j \, \bar f_j.
\end{align}
Comparing with
\begin{align}
    W_j \left(f_j - \bar f_j\right)^2 = W_j f_j^2 - 2 W_j f_j \bar f_j + W_j \bar f_j^2,
\end{align}
the two expressions differ only by $\sum_{i=1}^{I-j} w_{i,j} f_{i,j}^2 - W_j \bar f_j^2$, which does not involve $f_j$. The weighted loss and the compressed form $\sum_j W_j (f_j - \bar f_j)^2$ therefore share the same minimiser. The difference is an additive constant rather than a scale factor, so the balance against the penalty terms is unaffected and no scalar adjustment is required.

Defining the diagonal weight matrix $\mathrm{W} = \operatorname{diag}(W_1, \ldots, W_n)$, the compressed empirical loss is the quadratic form
\begin{align}
    \sum_{j=1}^{n} W_j \left(f_j - \bar f_j\right)^2
    = (\boldsymbol{f} - \bar{\boldsymbol{f}})^\top \mathrm{W} \, (\boldsymbol{f} - \bar{\boldsymbol{f}}).
\end{align}

\subsection{The penalties as quadratic forms}

The reference-pattern penalty of Equation~\eqref{eq:reference-penalty} is immediate:
\begin{align}
    J_{\text{ref}}(\boldsymbol{f})
    = \sum_{j=1}^{n} W_j \left(f_j - \dot f_j\right)^2
    = (\boldsymbol{f} - \dot{\boldsymbol{f}})^\top \mathrm{W} \, (\boldsymbol{f} - \dot{\boldsymbol{f}}).
\end{align}

The smoothness penalty couples the ratios, so it needs a difference matrix. Let $\mathrm{D}$ be the $(n-2) \times n$ second-difference matrix whose rows carry $(-1, 2, -1)$ on the appropriate shift, so that
\[
    (\mathrm{D}\boldsymbol{f})_j = 2 f_j - f_{j-1} - f_{j+1}, \qquad j = 2, \ldots, n-1,
\]
and let $\tilde{\mathrm{W}} = \operatorname{diag}(W_2, \ldots, W_{n-1})$. Then the smoothness penalty of Equation~\eqref{eq:smooth-penalty} is
\begin{align}
    J_{\text{smooth}}(\boldsymbol{f})
    = \sum_{j=2}^{n-1} W_j \left(2 f_j - f_{j-1} - f_{j+1}\right)^2
    = (\mathrm{D}\boldsymbol{f})^\top \tilde{\mathrm{W}} \, \mathrm{D}\boldsymbol{f}
    = \boldsymbol{f}^\top \mathrm{D}^\top \tilde{\mathrm{W}} \mathrm{D} \, \boldsymbol{f}.
\end{align}

\subsection{Normal equations and uniqueness}

The assembled core objective of Equation~\eqref{eq:assembled-objective}, in matrix form, is
\begin{align}
    \mathscr{L}(\boldsymbol{f})
    = (\boldsymbol{f} - \bar{\boldsymbol{f}})^\top \mathrm{W} (\boldsymbol{f} - \bar{\boldsymbol{f}})
    + \alpha_{\text{ref}} \, (\boldsymbol{f} - \dot{\boldsymbol{f}})^\top \mathrm{W} (\boldsymbol{f} - \dot{\boldsymbol{f}})
    + \alpha_{\text{smooth}} \, \boldsymbol{f}^\top \mathrm{D}^\top \tilde{\mathrm{W}} \mathrm{D} \boldsymbol{f},
\end{align}
with gradient
\begin{align}
    \frac{\partial \mathscr{L}(\boldsymbol{f})}{\partial \boldsymbol{f}}
    = 2 \mathrm{W} (\boldsymbol{f} - \bar{\boldsymbol{f}})
    + 2 \alpha_{\text{ref}} \mathrm{W} (\boldsymbol{f} - \dot{\boldsymbol{f}})
    + 2 \alpha_{\text{smooth}} \mathrm{D}^\top \tilde{\mathrm{W}} \mathrm{D} \boldsymbol{f}.
\end{align}
Setting the gradient to zero gives the normal equations
\begin{align}\label{eq:normal-equations}
    \left[(1 + \alpha_{\text{ref}}) \, \mathrm{W} + \alpha_{\text{smooth}} \, \mathrm{D}^\top \tilde{\mathrm{W}} \mathrm{D}\right] \hat{\boldsymbol{f}}
    = \mathrm{W} \left(\bar{\boldsymbol{f}} + \alpha_{\text{ref}} \, \dot{\boldsymbol{f}}\right).
\end{align}

The solution is unique: $\mathrm{W}$ is positive definite whenever every $W_j > 0$, which holds for any triangle with at least one observed ratio per development period and $\rho > 0$; $\mathrm{D}^\top \tilde{\mathrm{W}} \mathrm{D}$ is positive semi-definite; and $\alpha_{\text{ref}}, \alpha_{\text{smooth}} \ge 0$, so the system matrix is positive definite and the objective is strictly convex.

Two special cases confirm the construction. With $\alpha_{\text{smooth}} = 0$ the system matrix is diagonal and Equation~\eqref{eq:normal-equations} reduces per ratio to $(1 + \alpha_{\text{ref}}) W_j \hat f_j = W_j (\bar f_j + \alpha_{\text{ref}} \dot f_j)$ - the weights cancel and the credibility blend of Equation~\eqref{eq:credibility-blend} is recovered. With $\alpha_{\text{ref}} = \alpha_{\text{smooth}} = 0$ the solution is $\hat{\boldsymbol{f}} = \bar{\boldsymbol{f}}$, the weighted empirical fit.

\section{Gradient of the shift penalty}\label{ap:shift-gradient}

The shift penalty of Equation~\eqref{eq:shift-penalty} has no closed-form solution, so its gradient is the specification the optimiser runs on. The derivative is more involved than those of the shape penalties because the increments $h_m$ depend on the ratios recursively. Each $\pi_m$ is a product over all later ratios, so a single $f_j$ moves every increment at or before development period $j+1$. Throughout this appendix $m$ indexes increments and $j$ the ratio being differentiated.

\paragraph{Percentage developed.}
From $\pi_m = (f_m f_{m+1} \cdots f_n)^{-1}$, the ratio $f_j$ appears in $\pi_m$ exactly when $j \ge m$, and differentiating the product gives
\begin{align}\label{eq:dg}
    \frac{\partial \pi_m}{\partial f_j} = -\frac{\pi_m}{f_j} \, 1_{\{j \ge m\}}.
\end{align}

\paragraph{Incremental percentage developed.}
For $m > 1$, $h_m = \left(1 - f_{m-1}^{-1}\right) \pi_m$, and the product rule gives
\begin{align}
    \frac{\partial h_m}{\partial f_j}
    &= \pi_m \frac{\partial}{\partial f_j}\left(1 - f_{m-1}^{-1}\right)
     + \left(1 - f_{m-1}^{-1}\right) \frac{\partial \pi_m}{\partial f_j} \notag \\
    &= 1_{\{j = m-1\}} \, \frac{\pi_m}{f_{m-1}^2}
     - \left(1 - f_{m-1}^{-1}\right) \frac{\pi_m}{f_j} \, 1_{\{j \ge m\}} \notag \\
    &= 1_{\{j = m-1\}} \, \frac{\pi_m}{f_{m-1}^2}
     - \frac{h_m}{f_j} \, 1_{\{j \ge m\}}.
    \label{eq:dh}
\end{align}
The two indicators never fire together, so each ratio affects $h_m$ through exactly one channel: $f_{m-1}$ through the retained-share factor, and every $f_j$ with $j \ge m$ through the product $\pi_m$. For $m = 1$, $h_1 = \pi_1$ and Equation~\eqref{eq:dg} gives $\partial h_1 / \partial f_j = -h_1 / f_j$ for every $j$, consistent with Equation~\eqref{eq:dh} read at $m = 1$.

\paragraph{The slope.}
Because the regression weights $(m - \bar m)$ sum to zero over $m = 1, \ldots, n+1$, the mean term of the slope drops out and
\begin{align}
    \hat\beta(\boldsymbol{f}) = \frac{1}{S} \sum_{m=1}^{n+1} (m - \bar m) \, h_m,
    \qquad
    \bar m = \frac{n+2}{2},
    \qquad
    S = \sum_{m=1}^{n+1} (m - \bar m)^2 = \frac{n(n+1)(n+2)}{12},
\end{align}
where $S$ is a constant. Differentiating term by term with Equation~\eqref{eq:dh}, the indicator $1_{\{j = m-1\}}$ fires only at $m = j+1$ and the indicator $1_{\{j \ge m\}}$ collects the increments up to $m = j$:
\begin{align}\label{eq:dbeta}
    \frac{\partial \hat\beta}{\partial f_j}
    = \frac{1}{S} \left[ (j + 1 - \bar m) \, \frac{\pi_{j+1}}{f_j^2}
    - \frac{1}{f_j} \sum_{m=1}^{j} (m - \bar m) \, h_m \right].
\end{align}

\paragraph{The truncated slope.}
With $h_m^+ = \max(h_m, 0)$, an increment contributes to the gradient only where it is positive, since $\partial h_m^+ / \partial f_j = 1_{\{h_m > 0\}} \, \partial h_m / \partial f_j$ and $1_{\{h_m > 0\}} h_m = h_m^+$:
\begin{align}\label{eq:dbetastar}
    \frac{\partial \hat\beta^*}{\partial f_j}
    = \frac{1}{S} \left[ (j + 1 - \bar m) \, \frac{\pi_{j+1}}{f_j^2} \, 1_{\{h_{j+1} > 0\}}
    - \frac{1}{f_j} \sum_{m=1}^{j} (m - \bar m) \, h_m^+ \right].
\end{align}

\paragraph{The penalty.}
The chain rule through the sigmoid, using $\operatorname{sigmoid}'(x) = \operatorname{sigmoid}(x)\bigl(1 - \operatorname{sigmoid}(x)\bigr)$, gives the gradient the optimiser uses:
\begin{align}\label{eq:dshift}
    \frac{\partial J_{\text{shift}}}{\partial f_j}
    = \operatorname{sigmoid}\left(\hat\beta^*\right) \left(1 - \operatorname{sigmoid}\left(\hat\beta^*\right)\right)
      \frac{\partial \hat\beta^*}{\partial f_j},
\end{align}
with $\partial \hat\beta^* / \partial f_j$ the full recursive expression of Equation~\eqref{eq:dbetastar} - the collected sum over earlier increments must not be dropped, since it is the channel through which a ratio moves the whole tail of the profile.

\paragraph{Differentiability.}
The truncation makes $J_{\text{shift}}$ non-differentiable exactly where some $h_m = 0$. These are isolated points, so the penalty is differentiable almost everywhere. It is standard to treat it as differentiable in optimisation, at the cost of a rougher objective surface that favours robust gradient-based optimisers.

\end{document}